\documentclass[11pt]{article}
\usepackage{amssymb}
\usepackage{bbm}   
\usepackage{latexsym} 
\usepackage{epsfig}
\usepackage{color }
\usepackage{mathrsfs} 
\newtheorem{assump}{Assumption}
\newtheorem{theorem}{Theorem}[section]
\newtheorem{proposition}{Proposition}[section]
\newtheorem{corollary}[proposition]{Corollary}
\newtheorem{lemma}[proposition]{Lemma}
\newtheorem{definition}{Definition}[section]
\newtheorem{conjecture}{Conjecture}[section]
\newcommand{\Proofstart}{\textit{Proof: }}
\newcommand{\Proofend}{ $\square$ }
\def\bd{\begin{definition}}
\def\ed{\end{definition}}
\def\be{\begin{equation}}
\def\ee{\end{equation}}
\def\bea{\begin{eqnarray}}
\def\eea{\end{eqnarray}}
\def\bitem{\begin{itemize}}
\def\eitem{\end{itemize}}
\def\ben{\begin{enumerate}}
\def\een{\end{enumerate}}
\def\bdescribe{\begin{description}}
\def\edescribe{\end{description}}

\def\a{\alpha}
\def\b{\beta}
\def\etaperp{{\eta_\perp}}
\def\c{\gamma}
\def\d{\delta}
\def\e{\varepsilon}
\def\l{\lambda}

\def\r{\rho}
\def\s{\sigma}

\def\t{\tau}

\def\th{\theta}
\def\om{\omega}
\def\Om{\Omega}
\def\Omtilde{\tilde{\Omega}}
\def\Ups{\Upsilon}
\def\half{\frac{1}{2}}

\def\ap{{\a'}}
\def\bp{{\b'}}

\newcommand{\ca}{{\underline{\a}}}
\newcommand{\cb}{{\underline{\b}}}
\newcommand{\cc}{{\underline{\c}}}

\def\TD{{\mathcal{D}}} 
\newcommand{\JT}{N}
\newcommand{\JTtwo}{M}

\newcommand{\TCurv}{{\mathcal{R}}}
\newcommand{\CY}{Y}
\newcommand{\TE}{{\mathcal{E}}}
\newcommand{\Tg}{{\mathcal{G}}}
\newcommand{\TC}{{\mathcal{C}}}
\newcommand{\TB}{{\mathcal{B}}}
\newcommand{\TA}{{\mathcal{A}}}
\newcommand{\TAA}{A}
\newcommand{\TZ}{Z}
\newcommand{\cg}{\mathbf{g}}

\newcommand{\co}{\sigma}
\newcommand{\ct}{\tau}
\newcommand{\cu}{\mu}
\newcommand{\cp}{\rho}
\newcommand{\CV}{v}

\newcommand{\gtilde}{\tilde{g}}

\newcommand{\D}{\nabla}

\newcommand{\Dhat}{\hat{\nabla}}
\newcommand{\Dtilde}{\tilde{\nabla}}

\newcommand{\Phat}{\hat{P}}
\newcommand{\Rhat}{\hat{R}}
\newcommand{\Ptilde}{\tilde{P}} 

\newcommand{\btilde}{\tilde{b}}

\def\rest{\arrowvert} 
\def\implies{\Rightarrow}

\def\norm{\Vert}

\newcommand{\phm}{\phantom{-}}  
\def\sect{\in\Gamma}

\def\real{\mathbbm{R}} 

\def\Uset{{\mathcal{U}}}

\def\Wset{{\mathcal{W}}}

\newcommand{\tensor}[3]{_{#1 \phantom{#2}#3}^{\phantom{#1}#2}}

\newcommand{\vect}[3]{\left(\begin{array}{c}{#1}\\{#2}\\{#3}\end{array}\right)}

\def\non{\nonumber}

\begin{document}

\title{\textbf{An extension theorem \\[1ex] for conformal gauge singularities}}

\author{\\
\textbf{Christian L\"ubbe}
\\ \small{Mathematical Institute, Oxford OX1 3PN, United Kingdom}
\\ \small{Queen Mary University of London, London E1 4NS, United Kingdom}
\\
\\
\textbf{ Paul Tod}
\\ \small{Mathematical Institute, Oxford OX1 3PN, United Kingdom}
\\ \small{St John's College, Oxford OX1 3JP, United Kingdom}}
\maketitle

\begin{abstract}
We analyse conformal gauge, or isotropic, singularities in cosmological models in general relativity. Using the calculus of tractors, we find conditions in terms of tractor curvature for a local extension of the conformal structure through a cosmological singularity and prove a local extension theorem along a congruence of time-like conformal geodesics.
\end{abstract}

%
\section{Introduction}

A conformal gauge singularity is a singularity of the space-time metric which does not correspond to a singularity of the conformal metric. Rather, the singularity is wholly attributable to the choice of \emph{conformal gauge} or representative metric in the conformal class, and there will be other choices of representative metric with no singularity. The physical motivation for studying conformal gauge singularities, which have also been called isotropic singularities \cite{GW}, comes from cosmology. Since the Weyl tensor is conformally-invariant, it will necessarily be nonsingular, in a way to be made precise, at a conformal gauge singularity. 
Penrose \cite{PenroseEinsteinCentsurvey} has argued that, since the observed cosmic microwave background is so isotropic, the Big-Bang singularity was highly ordered and gravitational entropy was initially low. He formulated this idea geometrically as a conjecture that the Weyl tensor at the initial singularity should be finite, or even zero. The simplest formulation of this Weyl tensor conjecture (WTC) is that the initial singularity is an isotropic singularity, understood as a conformal gauge singularity. Isotropic singularities have been studied with the WTC in mind by Goode and Wainwright \cite{GW}, Newman \cite{NewmanIS}, and Anguige and Tod \cite{AT1,AT2}, among others. 

In this article we investigate the nature of singularities of the conformal structure, that is to say for the conformal metric rather than the metric. To see what difference this makes, consider the following situation: 
start with a manifold $M$ with a regular metric $g_{ij}$;  choose a smooth function $\Omtilde $ vanishing on a smooth space-like hypersurface $ \Sigma_0$ and define $\tilde{M} := \{ p \in M : \Omtilde(p)>0 \} $ and $\partial_0 M := \{ p \in M : \Omtilde(p)=0 \} $;
then the rescaled metric $\gtilde_{ij}=\Omtilde^2 g_{ij} $, ($\gtilde^{ij} = \Omtilde^{-2} g^{ij} $), is regular on $\tilde{M}$ but singular on $\partial_0 M$; however, the conformal class $[\gtilde]=[g] $ is regular in the sense that it has a regular representative. Thus the singularity in $(\tilde{M},\gtilde)$ is a result of the choice of representative metric in $[g]$, or equivalently of the choice of conformal gauge, but not of the conformal class itself: it is a \emph{conformal gauge singularity} (by analogy with a \emph{coordinate singularity}, which is due to choice of coordinates). 
This motivates the following definition:
\begin{definition}{\textbf{Conformal gauge singularity}}
\label{CGS2}

Let $(\tilde{M},\gtilde) $ be a space-time. Suppose there exists a manifold $M$ with boundary $\partial M$ such that
\ben
  \item $M = \tilde{M} \cup \partial M$ and for each $p \in \partial M$ there exists a conformal chart $(U, g, \Omtilde)$, with $U$ an open neighbourhood of $p$ in $M$ and $\Omtilde $ smooth on $U - \partial U$ such that
  \item $\partial U = U \cap \partial M $ and $U - \partial U = U \cap \tilde{M} $;
  \item $\gtilde_{ij} = \Omtilde^2 g_{ij}$ on $U - \partial U$;
  \item $\Omtilde[\partial U]=0$, $\Omtilde[U - \partial U] > 0$
\een
Then $ \partial M$ is a conformal gauge singularity of $(\tilde{M},\gtilde)$.
\end{definition}
A version of this definition given by Newman \cite{NewmanIS} contains the extra condition that $\tilde\Om$ be smooth with $d\tilde\Om$ nowhere-vanishing on $U$. 
Newman imposed this extra condition in order to use the conformal factor as a (smooth) cosmic time function. 
Here we do not require this, as there exist explicit examples of  conformal gauge singularities with conformal factors that are not smooth or have $d\Om=0$ at the boundary \cite{TodTorino}. 

We refer to $(\tilde{M},\gtilde)$ as the physical space-time  and to the conformally related space-time $(M,g)$ as the unphysical space-time. How can we distinguish a genuine singularity of the conformal class from one caused by the choice of conformal gauge? Furthermore how can we find a suitable rescaling $\Om =\Omtilde^{-1}$ to obtain the regular unphysical metric $g_{ij}$?

The usual approach to defining space-time singularities uses the concepts of extensions and of incompleteness of curves defined as follows. A manifold $(\tilde{M}, \gtilde)$ is said to admit an extension if there exists an embedding into a second manifold $(M,g)$ of the same dimension such that $g \rest_{\tilde{M}}=\gtilde$. 
A curve $\c$ in $\tilde{M}$ is said to be incomplete if it has a finite generalised affine parameter (g.a.p.) length (see e.g. \cite{HEbook,Clarkebook}) and no endpoint in $\tilde{M}$. Incompleteness may be due just to points missing from the space-time, which an extension may replace. If $\c$ remains inextendible in any extension of the space-time then there is a genuine singularity and it is often possible to define it in terms of ideal endpoints of curves like $\c$ \cite{Clarkebook,Schmidtbboundary}.

In our work we will follow these ideas and concepts with some suitable modifications. In order to accomodate for the conformal gauge singularities the definition of extensions needs to be generalised to the conformal metric. We will also need to adapt the definition of completeness as we do not have a natural generalised affine parameter for the conformal metric. Our analysis will make use of conformal geodesics and we will see that they have a fractional linear parameter freedom associated to them. Therefore it is ambigious to talk of finiteness of the conformal parameter in the context of completeness. Instead we will introduce the notion of \textit{infinitely far away} for conformal geodesics.

Since in general it will be difficult to find an explicit extension, it is more desirable to use geometric features of the space-time itself to characterise singularities and determine their nature. One may seek to classify space-times and singularities using some notion of strength or differentiability as was done by Ellis and Schmidt \cite{EllisSchmidtsingularST1,EllisSchmidtsingularST2} and Clarke \cite{Clarkebook,Clarke74,Clarkeclassification,Clarkenature,Clarkelowdiff}. 
It might be hoped that a sufficiently differentiable curvature tensor would indicate the absence of a singularity and the possibility of an extension. In this direction, both Clarke \cite{Clarkebook} and R{\'a}cz \cite{Racz} proved the existence of an extension of a certain differentiable order, given bounded curvature derivatives up to a certain other order. In this article we extend their work to the study of conformal metrics. 

For a given conformal class the Weyl curvature is conformally invariant in that $\tilde{C}\tensor{ij}{k}{l} = C\tensor{ij}{k}{l}$. 
Therefore a necessary condition for conformal gauge singularities is a finite Weyl tensor. We shall see below how to phrase this condition, and similar conditions on the derivatives of the Weyl tensor, in terms of components in a conformally-invariant way (see (\ref{curvnormfunctions})). Note  that the Ricci curvature changes under conformal rescaling and can thus be regular for the unphysical metric and singular for the physical metric. Thus it is unsuitable for recognising a conformal gauge singularity.

In order to characterise conformal gauge singularities completely, one also needs a set of sufficient conditions on the conformal structure. 
The aim of this article is to establish a theorem to the effect that, given a space-time singularity with suitable conditions of finiteness on the Weyl curvature and its derivatives, there is a metric in the conformal class which is nonsingular and can therefore be extended through the singularity. This is done by theorem \ref{Maintheorem}. The conditions of finiteness need to be formulated in a conformally-invariant way. We use conformal geodesics, Weyl propagation and tractors, all of which will be introduced below. Tractors define a conformally invariant calculus and have been used successfully to study conformally invariant differential operators \cite{BEG,CG1,CG2}.  We shall use them to analyse the curvature of the conformal structure in a gauge independent way. 

The plan of this article is as follows. In Section 2, we give some generalities on conformal geometry in dimension $n \ge 3$, including the theory of conformal geodesics and tractors, and a discussion of Jacobi fields and conjugate points for time-like conformal geodesics. This machinery is needed to be able to impose conformally-invariant curvature bounds which will then be necessary conditions for the existence of  a conformal extension. In Section 3, we restrict to dimension $n=4$ and quote the local extension  theorem of R\'acz for the space-time metric, and the extension theorem of Whitney for functions, which it is based on. Then we give the statement of our main result, Theorem \ref{Maintheorem}, a local extension theorem for the conformal metric. The proof of this needs a result on non-existence of conjugate points along time-like conformal geodesics given conditions on curvature, Theorem \ref{conjugate point theorem}. That section ends with Lemma \ref{coordinate system}, establishing Whitney's Property $\mathcal{{P}}$, a necessary condition for Whitney's Extension Theorem, for the sets we consider. In Section 4, with an intricate series of inductions, we complete the proof of Theorem \ref{Maintheorem}. The section also contains a global extension theorem, Theorem \ref{global extension theorem}, subject to some rather strong assumptions.

\section{Some conformal geometry}
\label{The conformal structure}
Here we review the conformal geometry necessary for our construction. We will recall some of the basic concepts related to conformal geometry and introduce the tractor formalism, which is well-known for the study of conformal and other parabolic geometries \cite{CG1}. The exposition follows closely to that in \cite{BEG}. However for readers with a background in general relativity, who are most likely unfamiliar with the details of this formalism, we will, for the sake of completeness, recall and highlight details that are of importance for the subsequent analysis and discussion. 

\subsection{Conventions}
Throughout we will use the abstract index notation of Penrose \cite{PR1}. Note since we follow \cite{BEG}, there are differences of detail from \cite{PR1}. Our curvature convention is $(\Dhat_i \Dhat_j - \Dhat_j \Dhat_i)v^k = \Rhat\tensor{ij}{k}{l}v^l$ and we use the $(+\,-\,\ldots\,-)$ signature, with space-time dimension $n \ge 3$ throughout this article. We will work with weighted tensor and tractors, defined below, and follow the conventions of \cite{BEG}. 
We adopt the following indexing conventions:
\bitem
\item $i,j,k \ldots \in \{0, \ldots, n-1\} $ - tensor indices;
\item $I,J,K \ldots \in \{0, \ldots, n+1\} $ - tractor indices;
\item $\a, \b, \c \dots \in \{0, \ldots, n-1\}  $ - indices in a frame;
\item $\ap, \bp, \c' \dots  \in \{1, \ldots, n-1\} $ - the space-like frame indices;
\item $0 \quad \quad$ - the time-like frame index ;
\item  $\ca, \cb,\cc \ldots \in \{0, \ldots, n-1\} $ - indices in a coordinate frame (distinguished from the other frames);
\item $\TB, \TC \dots \in \{0, \ldots, n+1\} $- indices in a tractor frame.
\eitem
Since $n$ of the $n+2$ components of a tractor are a vector, vector and tractor indices will be used in correspondence where appropriate i.e. $\a, \b, \c, \ldots$ with $\TA, \TB, \TC,\ldots$ respectively. When a frame tractor corresponds to a specific frame vector we use the same Greek label for the tractor e.g. $\TE_\TA = \TE_\a $. For a bundle $B$ over the manifold $M$, $\Gamma(B) $ denotes the space of sections.

When discussing ODEs, a subscript $*$ on a quantity will indicate an initial value, e.g. $v_* $ denotes the initial velocity.

\subsection[The conformal metric and general Weyl connections]{The conformal metric and general Weyl connections}

For the purpose of this paper the metrics have signature $(1,n-1)$, but all definitions generalise to a signature $(p,n-p) $.
Two metrics $g$ and $\tilde{g}$ are said to be conformally related if there exists a smooth positive function $\Omega$ such that $g_{ij} = \Omega^2 \tilde{g}_{ij} $. Metrics conformally related to a chosen metric $g$ form the conformal class $[g]$, which can be viewed as a smooth subbundle of $S^2 T^*M$ with fibres $\real^+ $. Given a conformal class there is a whole family of associated real line bundles $\e[w]$ whose elements under the rescaling of the metric change like $q=\Omega^\omega \tilde{q} $. These bundles are referred to as conformally weighted line bundles $\e[w] $, where $w \in \real$ is the \emph{conformal weight}. 
Two sections can be added when they have the same conformal weight $w$ and their sum will have the same weight $w$. The product of two sections of weight $w_1 $ and $w_2 $, respectively, has weight $w_1 + w_2 $. 
A non-vanishing $\s\sect(\e[1]) $ is referred to as a \emph{conformal scale} and we can define its inverse $\s^{-1}\sect(\e[-1])  $, so that $\s\s^{-1} = 1 $, as well as powers $\s^k\sect(\e[k])  $. The bundle of scalars $\e$ can be identified with $\e[0] $.  

A conformal scale $\s$ defines a connection $\D $ on $\e[w] $ by
\be
\label{Ewconn1}
  \D_a \t = \s^{w} \partial_a (\s^{-w}\t).
\ee
This implies $\D_a \s =0 $, since $\s$ is non-vanishing. Using $\t= (\s^{-w}\t) \s^w = T \s^w$ we can write (\ref{Ewconn1}) for a non-vanishing section $\t$ as $\D_a \t = \frac{\partial_a T}{T} \t $. So for a connection on $\e[w] $, as defined in (\ref{Ewconn1}), the derivative of a non-vanishing section $\t$ is a gradient times $\t$ itself.

Any other non-vanishing section $\tilde{\s} \sect(\e[1]) $ can be written $\tilde{\s} = \Om^{-1}\s$, where $\Om $ is a nowhere vanishing function. For $\tilde{\s}$ we define the connection $\Dtilde$ analogously to (\ref{Ewconn1}). Setting $\Ups_a = \frac{\partial_a \Om}{\Om} $, it follows that for $\t\sect(\e[w]) $ the connections $\D$ and $\Dtilde$ are related as follows
\be
\label{sectiontrans}
  \Dtilde_a \t = \D_a \t + w \Ups_a \t .
\ee
We may define a more general connection on $\e[w] $ as follows: choose $\s\sect(\e[1]) $ and a 1-form $b_a $ and define $$\Dhat_a \s = b_a \s .$$ For $\t \sect(\e[w])$ the identity $\t = \s^w \s^{-w}\t $ and the Leibniz rule give
\be
\label{Ewconn2}
  \Dhat_a \t = \s^{w} \partial_a (\s^{-w}\t) + w b_a \t .
\ee
This gives a generalisation of  (\ref{Ewconn1}). For $\tilde{\s} = \Om^{-1}\s$ it follows that $\Dhat_a \tilde{\s} = \btilde_a  \tilde{\s} $, where $\btilde_a=b_a-\Ups_a $. Hence the choice $\tilde{\s}$ with 1-form $\btilde_a$ defines the same connection $\Dhat$. By (\ref{Ewconn2}) the connections are related as follows
\be
\label{Weylsectiontrans}
  \Dhat_a \t = \D_a \t + w b_a \t = \Dtilde_a \t + w \btilde_a \t
\ee 
Following Friedrich \cite{Friedrich}, we call (\ref{Weylsectiontrans})   a \emph{connection translation} and write it schematically as $\Dhat = \D + b = \Dtilde + \btilde$. Evidently, there exists a non-vanishing section $\hat{\s} \sect(\e[1])$ preserved by such a connection if and only if $b_a$ is exact.

Given a bundle $E$ over $M$ with connection $\D^E$, one can construct an associated conformally weighted bundle $E[w]  =E \otimes \e[w]  $ with a connection  $\Dhat^E$ acting on ${\cal{T}} \sect(E[w]) $, where $\s, b_a $ and $\Dhat$ as in (\ref{Ewconn2}),  by
\be
\label{Ewgeneralconn}
  \Dhat^E_a {\cal{T}} = \s^{w} \D^E_a (\s^{-w}{\cal{T}}) + w b_a {\cal{T}} .
\ee
It is easy to see that for weight-zero sections of $E$, (\ref{Ewgeneralconn}) reduces to $\D^E $ and for scalar densities to (\ref{Ewconn2}). Given a choice of metric, 1-form and conformal scale one can thus define a connection on weighted tensors using (\ref{Ewgeneralconn}). 

For a canonical definition of the connection one typically identifies the volume density bundle $\e[n]_{[\mu_1 \ldots \mu_n]} $ of the conformal class $[g]$ via $\vert g \vert \equiv \vert \mathrm{det}(g_{\ca\cb})\vert$ with $\e[n] $. This then associates a density $\s\sect(\e[1])$ canonically to the chosen metric $g_{ij} $. Since the Levi-Civita connection $\D$ of $g_{ij} $ preserves the associated volume form the identification induces a connection on $\e[w]$ which preserves $\s$ and coincides with the connection defined in (\ref{Ewconn1}).

One defines the conformal metric and its inverse as
\be
\label{conformalmetric}
  \cg_{ij}:=\s^2 g_{ij}, \quad \cg^{ij}:=\s^{-2} g^{ij}.
\ee
These are sections of $\e_{(ij)}[2] $ and $\e^{(ij)}[-2]$ respectively.  A conformal metric is hence a global symmetric non-degenerate tensor field  $\cg_{ij}$ with values in the line bundle $\e[2]$. It follows from (\ref{conformalmetric}) that
\be
\label{Dcg}
  \D_i \cg_{jk} = 0 ,\quad \D_i \cg^{jk} = 0.
\ee
If we choose the pair $(\tilde{\s}, \gtilde_{ij})= (\Om^{-1}\s, \Om^2 g_{ij})$ then this defines the same conformal metric and a new connection $ \Dtilde$. We take the view point that a choice of conformal scale $\s$ defines a representative metric $g_{ij}=\s^{-2}\cg_{ij}$ for the conformal class $[g]$ and a connection $\D$ that preserves $\cg_{ij} $ as well as $\s$ and $ g_{ij} $.

If $\Dhat_a$ is any torsion-free connection satisfying (\ref{Dcg}) then, with $b_a := \frac{\Dhat_a \co}{\co}$, we have
\be
\label{Dhatg}
 \Dhat_a g_{jk} = -2 b_a g_{jk}, \quad \Dhat_a g^{jk} = 2 b_a g^{jk}. 
\ee
These torsion-free connections are the \emph{Weyl connections} associated to $[g]$. It follows from our earlier statement and the above construction that a torsion-free connection $\Dhat$ preserving the conformal metric $\cg_{ij} $ is the Levi-Civita connection of a metric in the conformal class if and only if there exists a section $\hat{\s} \sect (\e[1])$ which is preserved by $\Dhat$.

We define the conformally invariant tensor 
\be
\label{Stensor}
S\tensor{ij}{kl}{}=\d\tensor{i}{k}{} \d\tensor{j}{l}{} + \d\tensor{i}{l}{}\d\tensor{j}{k}{} - \cg_{ij}\cg^{kl},
\ee
which is symmetric in both pairs of indices and covariantly constant in all Weyl connections. Then this relates the Christoffel symbols of the connections $\Dhat $ and $ \D $:
\be
\label{connectionchange} 
  \hat{\Gamma}^{k}_{ij}=  \Gamma^{k}_{ij} + S\tensor{ij}{kl}{} b_l.
\ee
By applying the Leibniz rule to $\e^i[w] = \e^i \otimes \e[w]$ and $\e_i[w] = \e_i \otimes \e[w]$ and (\ref{connectionchange}) we get
\bea
\label{tensorsectiontransform}
  \Dhat_i U^k &=& \D_i U^k + S\tensor{ij}{kl}{} b_l U^j + w b_i U^k , \non \\
  \Dhat_i \omega_j &=&\D_i \omega_j - S\tensor{ij}{kl}{}b_l \omega_k + w b_i \omega_j  .
\eea
These can be generalised to any type of tensors by further applications of the Leibniz rule. Setting $w=0$ we recover the transformation rule for weight-zero tensors, as given in \cite{Friedrich}. 

Above we have seen that there is a correspondence between sections $\co$ and Levi-Civita connections $\D$. Similarly there exists a $1-1 $ correspondence between the 1-forms $b$  and Weyl connections $\Dhat$. In this article all connections will be Weyl connections associated to the conformal metric $\cg$. For notational purposes we will denote a general Weyl connection by $\Dhat$, whereas if conformal scales $\s$ and $\tilde{\s}$ respectively metrics $g$ and $\gtilde$ in $[g]$ have been specified, $\D$ and $\Dtilde$ will denote their Levi-Civita connection. For reasons that will become clearer in the context of tractors, a choice of connection will be referred to as choice of \emph{conformal gauge}. 

The curvature for the connection $\Dhat$ can be decomposed as follows
\be
\label{curvdecomp}
\Rhat\tensor{ij}{k}{l} = 2 \{\d\tensor{}{k}{[i}\Phat_{j]l} - \cg_{l[i}\Phat_{j]m}\cg^{mk} - \Phat_{[ij]}\d\tensor{}{k}{l}  \} +C\tensor{ij}{k}{l},
\ee
where the Weyl tensor $C\tensor{ij}{k}{l}$ is the trace-free part of the Riemann tensor, while $\Phat_{ij}$ is the Schouten or Rho-tensor given in terms of the Ricci tensor, recall $n \ge 3$, by
\be
\label{Rhotensordef}
\Phat_{ij} = \frac{1}{n-2} \left( \Rhat_{(ij)} + \frac{n-2}{n}\Rhat_{[ij]} - \frac{1}{2(n-1)}\cg_{ij}\cg^{mn}\Rhat_{mn}\right).
\ee
Note that the Riemann, Weyl and Schouten tensors for $\Dhat$ have weight zero and the decomposition is independent of the choice of metric in $[g]$. The Weyl tensor is conformally invariant, whereas under (\ref{connectionchange}) the Schouten tensor transforms as
\be 
\label{Schouten}
\ P_{ij} - \Phat_{ij} = \D_i b_j - \half b_k b_l S \tensor{ij}{kl}{} .
\ee
For a Levi-Civita $\D$, $P_{ij}$ is symmetric and hence both $\Phat_{ij}$ and $\Rhat_{ij}$ will be symmetric if and only if $b$ is closed, i.e. locally arises from a conformal rescaling. 
Thus for a general Weyl connection the  term $\Phat_{[ij]}\d\tensor{}{k}{l}$, which vanishes for metric connections, breaks the anti-symmetry in the second pair of indices in $\Rhat_{ijkl} = \Rhat\tensor{ij}{m}{l}\cg_{mk} \sect(\e_{ijkl}[2]) $. 

To interpret $\Phat_{[ij]}$, use (\ref{Weylsectiontrans}, \ref{tensorsectiontransform}, \ref{Schouten}) and the symmetry of $S\tensor{ij}{kl}{}$ to obtain
\be
\label{dco}
  \Dhat_{[i} \Dhat_{j]} \tau = w \Dhat_{[i} b_{j]} \tau = - w \tau \Phat_{[ij]}.
\ee
Thus $\Phat_{[ij]}$ is the curvature of the connection on weight-one functions. The right hand side vanishes if and only if $w=0$ or $\Dhat$ is locally a metric connection. Although (\ref{dco}) arises from $\Dhat=\D + b$, with $\D_i \co=0$, the term $\D_{[i}b_{j]}=-\Phat_{[ij]}$ is invariant under change of the connection or $b \to b - \Ups$.  

The result (\ref{dco}) extends to weighted tensors by application of the Leibniz rule. For example for $\cu^k\sect(\e^k[w])$ we get
\be
\label{dcu}
  (\Dhat_{i}\Dhat_{j}-\Dhat_{j}\Dhat_{i})\cu^k = (\Rhat\tensor{ij}{k}{l} - 2w \Phat_{[ij]}\d\tensor{}{k}{l}) \cu^l .
\ee
Note for $w=-1$, $\Rhat_{ijkl} + 2 \Phat_{[ij]}\cg_{kl} = C_{ijkl} + 2 \cg_{k[i}\Phat_{j]l} - 2 \cg_{l[i}\Phat_{j]k}$ is anti-symmetric in $kl$.
One can see that any higher derivative of a weighted tensor may have extra $\Phat_{[ij]}$ terms in it. Since \cite{BEG} only consider metric connections these terms vanish from their formulae. 

For a general Weyl connection and $n > 3$ the Bianchi identity and its contracted form are given by
\bea
\label{WeylBianchi}
  2(n-3) \Dhat_{[i}\hat{P}_{j]l} = \Dhat_k C\tensor{ij}{k}{l} \\
  \implies g^{il}\Dhat_i \hat{P}_{jl} = g^{il}\Dhat_j \hat{P}_{il} = \Dhat_j \hat{P} - 2 b_j \hat{P} , \non
\eea
where $\Dhat = \D + b$ and $\hat{P}=\hat{P}_{il}g^{il}$. The tensor $\widehat{\CY}_{ijk} = 2\Dhat_{[i}\hat{P}_{j]l}$ is known as the Cotton-York tensor \cite{BEG}. Equation (\ref{WeylBianchi}) also holds for $n=3$, since the Weyl tensor vanishes identically in that case. From (\ref{tensorsectiontransform}) we get 
\be
\label{divWeyl}
\Dhat_k C\tensor{ij}{k}{l}=\D_k C\tensor{ij}{k}{l} + (n-3) b_k C\tensor{ij}{k}{l}
\ee
and thus for $ \D + b= \Dhat = \tilde{\D} + \tilde{b} $ we can rewrite (\ref{divWeyl}) as
\be
\label{CYtransform}
  \CY_{ijl} + b_k C\tensor{ij}{k}{l}= \widehat{\CY}_{ijl} =  \tilde{\CY}_{ijl} + \tilde{b}_k C\tensor{ij}{k}{l}.
\ee

\subsection{Tractors}

The tractor formalism is a conformally invariant calculus which we shall use to analyse the singularity. 
We follow the setup of \cite{BEG} and the definitions are reviewed below.
\bd
  For a conformal manifold $(M, \cg)$ with a choice of associated general Weyl connection $ \D$ we introduce the direct sum 
\be
  \e^I = \e[1] \oplus (TM \otimes \e[-1]) \oplus\e[-1]  . \non
\ee
and for a connection translation $\Dhat = \D + b$  the different sections are identified according to 
\be
\label{tractortrans}
  \vect{\hat{\co}}{\hat{\cu}^i}{\hat{\cp}}=\vect {\co} {\cu^i+b^i\co} {\cp-b_j\cu^j-\half b_j b^j\co}_{.}
\ee
where $\co, \hat{\co}\sect(\e[1]); \cu^i, \hat{\cu}^i \sect(\e^i[-1]); \cp, \hat{\cp}\sect(\e[-1])$. Then $\e^I$ is the tractor bundle of $(M,\cg)$ .
\ed

A tractor index is denoted by a capital letter, with the corresponding section of $TM \otimes \e[-1]$ being identified with the same lower-case letter. By definition each general Weyl connection defines a unique splitting of the tractor bundle.  Thus the tractor bundle is invariant under change of conformal gauge, i.e. under connection translations (\ref{Weylsectiontrans}).

The sections $\co, \cu^i, \cp$ are ranked by the maximal power of $b$ in (\ref{tractortrans}) and refered to as the primary, secondary and tertiary parts of the tractor. We can see that $\co$ is conformally invariant. More generally if all higher ranked parts of a tractor vanish then the first non-zero part, called the projecting part, is conformally invariant. This generalises to tractors of any valence. For more detail the reader is referred to the discussion on composition series in \cite{BEG}.

\bd
The tractor metric $\Tg_{IJ} \sect(\e_{IJ}[0])$ is the non-degenerate symmetric form of signature $(p+1,q+1)$ given in block form by:
\be
\label{tractormetric}
  \left( \begin{array}{ccc} 0 & 0 & 1\\ 0 & \cg_{ij} & 0 \\ 1 & 0 & 0 \end{array} \right)_{.}  \non
\ee
Thus for $U^I = \vect{\co}{\cu^i}{\cp}$, $V^J = \vect{\varsigma}{\nu^j}{\varrho}$ we have $\Tg_{IJ}U^I V^J = \co \varrho+ \cg_{ij}\cu^i\nu^j +\cp\varsigma$. 

\ed
The definition is conformally invariant. The tractor metric $\Tg_{IJ}$ also gives an isomorphism between $\e^I$ and its dual $\e_I = \e[-1] \oplus (T^*M \otimes \e[1]) \oplus\e[1]  $, which is used to raise and lower tractor indices e.g. $U_I=(\cp, \, \cu_i, \, \co)$. We define the bundle of weighted tractor sections as $\e^I[w]=\e^I \otimes \e[w]$. All dual tractors of the form $P_I=\left(\cp, \, 0, \, 0 \right)$ define a subspace isomorphic to $\e[-1]$ and have a gauge invariant representation. We can see that the particular tractor $X_I:=\cp^{-1} P_I \sect(\e_I[1])$ gives an invariant injection $\e[-1] \to \e_I$ as well as a projection of $U^I$ onto its primary part $\co=U^I X_I$.

\bd
Given a Weyl connection $\Dhat$, the tractor connection $\TD$ on $\e^I$ is defined by
\bea
\label{tractorconnection}
  \TD_i \vect{\co} {\cu^j} {\cp}&=&\vect{\Dhat_i\co-\cu_i} {\Dhat_i\cu^j + \Phat\tensor{i}{j}{}\co + \d^j_i\cp } {\Dhat_i\cp-\Phat_{ik}\cu^k}_{.} 
\eea
\ed
Combining (\ref{tensorsectiontransform}) and (\ref{tractortrans}) one can see that the definition is gauge independent. $\TD$ preserves the tractor metric, i.e. $\TD_i \Tg_{JK}=0$, and hence raising and lowering commute with $\TD$. 
We can use a coupled Levi-Civita-tractor connection to define the tractor $\d_{iJ}=\TD_i X_J \sect(\e_{iJ}[1])$, so that $\d\tensor{}{i}{J} \sect(\e\tensor{}{i}{J}[-1] )$ always takes the form $(0, \, \d\tensor{}{i}{j}, \, 0 )$. It isolates the secondary part $\d\tensor{}{i}{J}U^J=\cu^i $ of a tractor in the chosen gauge and changes it accordingly under (\ref{tractortrans}). 

\bd
  The tractor curvature $\TCurv$ of $\TD$ acting on $\e^I$ is defined by
\be
  \left( \TD_i \TD_j - \TD_j \TD_i \right)U^K = \TCurv\tensor{ij}{K}{L} U^L .
\ee
\ed
On lower tractor indices therefore $$ \left( \TD_i \TD_j - \TD_j \TD_i \right)U_L = - \TCurv\tensor{ij}{K}{L} U_K$$ and $$\TCurv\tensor{ij}{M}{L}\Tg_{MK} = \TCurv_{ijKL}=\TCurv_{[ij][KL]}\mbox{  and  } \TCurv_{ijKL} X^L = 0.$$ These relations are best seen by calculating the tractor curvature in matrix form:
\be
\label{tractorcurvmatrix}
\left( \TD_i \TD_j - \TD_j \TD_i \right) \vect{\co}{\cu^k}{\cp} =
\left( \begin{array}{ccc}
 0 & 0 & 0\\
 2\Dhat_{[i}\Phat\tensor{j]}{k}{} & C\tensor{ij}{k}{l} & 0 \\
 0 & -2\Dhat_{[i}\Phat\tensor{j]}{}{l} & 0
 \end{array} \right) \vect{\co}{\cu^l}{\cp}_{.}
\ee
 This is the decomposition given in \cite{BEG} and it is independent of the connection (though note that this is still true with a Weyl connection, as here). For $n>3$ the identity (\ref{CYtransform}) can be recovered from (\ref{tractorcurvmatrix}) by applying (\ref{tractortrans}). For $n=3$ the Weyl tensor vanishes identically and $\hat{Y}_{ijl} $ is the conformally invariant projecting part, which agrees with (\ref{CYtransform}) . We can see from (\ref{WeylBianchi}) that for $n>3$ the tractor curvature consists of the Weyl tensor and its contracted derivative. Thus expressions involving derivatives of the tractor curvature depend only on the Weyl curvature and its derivatives, making it a suitable measure of the conformal structure. The tractor gauge will determine the connection used for the Weyl tensor derivatives.

Let $\Gamma$ be a congruence of smooth time-like curves $\c(\t)$ in $(M,\cg)$ with $v^i \partial _i \t =1$, where $v^i$ denotes the velocity vector field. Then $\CV=\sqrt{\cg(v,v)} \sect(\e[1])$ defines a canonical nowhere vanishing section associated to $\Gamma$. For a space-like congruence we choose $\CV=\sqrt{-\cg(v,v)} $. In this scale the velocity is automatically a unit vector and we refer to the associated Levi-Civita connection $\D$ as the unit velocity gauge.
We define the unit velocity section $u^i := \frac{v^i}{\CV}\sect(\e^i[-1])$ and the normed acceleration $\hat{a}^i:=\Dhat_u u^i\sect(\e^i[-2])$ with respect to the connection $\Dhat$:

For differentiation of tractors along the curves of the congruence we use $\TD_v \equiv v^i\TD_i$. We associate the tractor $Z^I = \CV^{-1}X^I$ to the congruence and define the associated velocity and acceleration tractors by
\be
\label{curvetractor}
   V^I = \TD_v Z^I, \qquad   A^I = \TD_v V^I .
\ee
In the $\Dhat$-gauge $V^I$ and $A^I$ take the form
\be
\label{VAform}
 V^I =  \vect{0}{u^i}{ - \langle b,u \rangle}, \quad 
A^I=\vect{-\CV}{\CV [\hat{a}^i - u^i \langle b,u\rangle ]  }{-\CV [\Dhat_u \langle b,u \rangle + \Phat(u,u)]}_{,}
\ee
where $\frac{\Dhat_i \CV}{\CV}=b_i$ gives $\Dhat= \D + b$. We observe that
\be
\label{tractorVAZ}
  V^I V_I=1, \quad V^I A_I=0 , \quad Z^I Z_I = 0, \quad Z^I V_I = 0, \quad Z^I A_I = -1.
\ee
The condition $A^I A_I=0$ fixes the parameterisation of the curve up to fractional linear transformations, see \cite{BEG} Prop. 2.11.

\subsection{Conformal geodesics}

In this subsection, we define conformal geodesics, and discuss their expression in terms of tractors and the interpretation of them given by the Schmidt gauge. The results of this are summarised in Proposition 2.1. We then discuss a potential problem with conformal geodesics, that the parameter freedom leads to quantities blowing-up at regular points of the manifold.

A conformal geodesic is a curve with tangent $v^i$ for which there exists a general Weyl connection such that 
\be
\label{Weylcgdef}
  \Dhat_v v^i=0 , \quad \quad\Phat_{ij}v^i=0 .
\ee
After a connection translation $\Dhat = \D + b$ these equations take the form
\bea
\label{Cgequ1}
  \D_v v^k + b_l S\tensor{ij}{kl}{}v^i v^j &=& 0 , \\
\label{Cgequ2}
  \D_v b_i - \half b_k b_l S\tensor{ij}{kl}{}v^i &=& P_{ij}v^i .
\eea
Under connection translation $\D \to \Dtilde=\D + \tilde{b}$ the conformal geodesic remains a conformal geodesic and only its $1$-form changes as $b \to b-\tilde{b}$. Choosing a scale $\co$ with $\D_i\co=0$, the 1-form $b$ associated to the conformal geodesic coincides with the one used in (\ref{Weylsectiontrans}). We define the conformal parameter $\t$ by $\D_v \t = 1$. Note that $v^i$ and $\t$ are invariants of the conformal class, but that fractional linear transformations of $\t$ preserve the conformal geodesic as a point set \cite{BEG}. We shall see this below. Null conformal geodesics are actually null geodesics and a further invariant of the conformal class.  Leaving these aside for the moment we have:

\begin{proposition}
The following conditions are equivalent definitions for a time-like or space-like conformal geodesic $\gamma$ associated to $\cg$
\bitem
\item There exists a general Weyl connection for $\cg$ such that the velocity satisfies 
\be
  \Dhat_v v^i=0 , \quad \quad
  \Phat_{ij}v^i=0 . \non
\ee
\item
There exists a 1-form $b_i$ along $\c$ satisfying
\be
  \D_v v^k + b_l S\tensor{ij}{kl}{}v^i v^j = 0  , \quad \quad
  \D_v b_i - \half b_k b_l S\tensor{ij}{kl}{}v^i = P_{ij}v^i . \non
\ee
\item
The acceleration tractor $A^I$ of $\c$ satisfies
 \be \TD_v A^I=0 , \quad \quad A^I A_I=0 . \non \ee
\item
Locally there exists a conformal rescaling to a metric $\tilde{g}$ for which $\c$ becomes a metric geodesic and $\Ptilde_{ij}v^i$ vanishes identically along $\c$.
\eitem
\end{proposition}

The tractor definition does not work for null conformal geodesics, whereas all the others do.

\Proofstart
We've seen the first two already. For the third, combining (\ref{curvetractor}) in the $\Dhat$-gauge with (\ref{Weylsectiontrans}),  (\ref{Weylcgdef}), $\frac{\Dhat_i \CV}{\CV}=b_i$ and $\langle b,v \rangle=0$ we see that time-like or space-like conformal geodesics satisfy
\be
\label{cgtractor}
  V^I=\vect{0}{u^i}{0}, \quad A^I=\vect{-\CV}{0}{0}, \quad \TD_v A^I=0, \quad A^I A_I=0 .
\ee
Using (\ref{tractortrans}) with $\tilde{b}$ we get 
\be 
V^I=\vect{0}{u^i}{\langle \tilde{b},u \rangle}, \quad  A^I=\CV \vect{-1}{ \tilde{b}_i}{\half \tilde{b}_k \tilde{b}^k} .
\ee
Thus for a conformal geodesic $A^I $ depends only on the gauge-dependent 1-form and the scalar velocity section. So $b_i$ plays the role of an acceleration. This is best seen in the unit velocity gauge where (\ref{Cgequ1}) takes the form $\D_v v^i=b^i$ as $\langle b,v \rangle = 0 $. 

\noindent Conversely, as can be seen in the $\Dhat$-gauge, $\TD_v A^I=0$ and $A^I A_I=0$ imply that the curve satisfies (\ref{Weylcgdef}) and, as stated earlier, the nullness of $A$ leads to the freedom to perform fractional linear or M\"obius transformations in $\tau$ \cite{BEG}. We observe that $\Phat_{ij}v^i=0$ implies that, when using $\Dhat$ for any differentiation along the conformal geodesic, the Schouten tensor terms vanish from the expression (\ref{tractorconnection}).

For the fourth result, we isolate a single conformal geodesic $\c$ and seek a function $\Omega$ in a neighbourhood of this curve which, on this curve, satisfies
\bea
  \D_i\Om &=& \Om b_i \, ,\non \\
\label{Schmidt gauge}
  v^i \D_i \D_j \Om &=& \Om v^i(P_{ij} +2 b_i b_j - \half \cg_{ij}\cg^{kl}b_k b_l) .
\eea
For any starting point $\c(\t_0) $ this system has a solution on an interval $I = [\t_0 - \l, \t_0 + \l] $. Rescaling with $\Omega$, in the new gauge the velocity is a unit vector and the new $\tilde{b}_i$ and $\Ptilde_{ij} $ vanish on $\c[I]$. Thus in the new gauge $\c[I]$ is a metric geodesic segment, and this leads to an alternative definition given by Schmidt (\cite{Schmidt}): a conformal geodesic is a curve $\c$ such that locally there is a conformal rescaling, which we shall call the `Schmidt gauge', for which $\c[I]$ becomes a metric geodesic and $\Ptilde_{ij}$ vanishes identically along $\c[I]$. \Proofend

\noindent{\bf{NB}} For the remainder of this article we shall only be concerned with time-like conformal geodesics. Results similar to those we shall find can be obtained using null (conformal) geodesics but will be the subject of a later paper \cite{Luebbenullextthm}.

In the Schmidt gauge, if the segment on which $\Om$ is defined reaches a singularity at which the Weyl tensor is finite, then in the rescaled space-time we have an incomplete metric geodesic with finite Riemann curvature even at the singularity and, following Clarke \cite{Clarke74} and R{\'a}cz \cite{Racz} we expect an extension to be possible. This motivates the use of conformal geodesics to produce conformal extensions. 

We need to look further at existence of conformal geodesics, and a difficulty introduced by the freedom to perform fractional linear transformations in the preferred parameter. Because of this freedom, $b_i$ can have poles and $v^i$ zeroes at regular points of the manifold. To see how this works, 
choose a scale $\co$ and define $q^2:=\frac{\CV}{\co}$. Then we can write 
\be\label{rexc}
v^i=q^2 u^i,
\ee
where $u^i$ is the unit velocity related to the scale $\co$. Let $\Dtilde$ denote the Levi-Civita connection associated to $\co$, i.e. $\Dtilde_i \co = 0$. The proper time related to $\gtilde_{ij}=\co^{-2}\cg_{ij}$ is denoted by $t$ and we set $D=\Dtilde_u=\frac{d}{dt}$. Rewriting (\ref{Cgequ1}, \ref{Cgequ2}) in terms of $q, u^i, a^i= D u^i$ gives
\bea 
\label{ucg1}
  Da^i - \gtilde(Da,u)u^i &=& \tilde{P}\tensor{j}{i}{}u^j - \tilde{P}(u,u)u^i \\
\label{ucgq}
  -2\frac{D^2 q}{q} &=& \tilde{P}(u,u) + \half \gtilde(a,a) , \\
  \label{ucgb}
  \tilde{b}_i &=& a_i - 2\frac{Dq}{q} u_i .
\eea
\noindent Alternatively we can recover (\ref{ucg1}-\ref{ucgb}) by combining (\ref{curvetractor}, \ref{VAform}, \ref{cgtractor}). 

Equation (\ref{ucg1}) and the initial data $(p_* \in M, u^i_*, a^i_*) $ determine the conformal geodesic as a point set. As long as the connection coefficients $\Gamma^k_{ij} $ and the Schouten tensor $P_{ij} $ are smooth, the curve $\c(t)$ is well defined with smooth unit velocity $u^i$ and acceleration $a^i$.

Next, (\ref{ucgq}) has a 2-parameter family of solutions depending on the initial values $(D q)_*, q_* $. 
In the Schmidt gauge $a^i$ and $P_{ij}u^i$ vanish along the curve, so from (\ref{ucgq}) it follows that $q$ will be a linear function of the related proper time depending on the initial data $(D q)_*, q_* $. Thus the function $q(t)$ can vanish at a regular point $p \in \c(t)$, then $v^i$ will have a zero there by (\ref{rexc}) and, by (\ref{ucgb}), $b_i$ a pole.  Furthermore, from (\ref{ucgb}), these are the only singularities which $b_i$ can have at points where the metric is regular.

From (\ref{rexc}), each solution of (\ref{ucgq}) corresponds to a different conformal parameter
\be
\label{definetau}
\t = \int \frac{dt}{q^2} \,.
\ee
Choose solutions $q_1$, $q_2$ of (\ref{ucgq}) with unit Wronskian:
$$q_1Dq_2-q_2Dq_1=1$$
then the general solution is $q=aq_1+bq_2$. From the unit Wronskian, we have
$$D\left(\frac{q_2}{q_1}\right)=\frac{1}{q_1^2}$$
so that
$$d\t=\frac{dt}{q^2}=\frac{dt}{q_1^2}(a+b\frac{q_2}{q_1})^{-2}$$
so if we introduce $\t_1$ satisfying
$$d\t_1=\frac{dt}{q_1^2}$$
then
$$d\left(\frac{q_2}{q_1}\right)=\frac{dt}{q_1^2}=d\t_1$$
and so
$$d\t=\frac{d\t_1}{(A+B\t_1)^2}$$
for some constants $A$ and $B$.
This is the freedom to change the conformal parameter $\t$ by a fractional linear transformation, \cite{BEG}, \cite{Friedrich}. 

The conformal geodesic can be expressed in terms of $q, a_i, u^i$ and the proper time $t$ using (\ref{ucg1}-\ref{ucgb}) or in terms of $v^i, b_i, \t$ using (\ref{Cgequ1}, \ref{Cgequ2}). The latter formalism is conformally-invariant but can generate singular behaviour due to gauge, where the first formalism remains regular. 

Suppose $\gtilde_{ij}= \co^{-2} \cg_{ij} $ is the physical metric and $ \tilde{b}_i$ is the 1-form associated to $\c$ in this gauge. Rescaling by $\Om = q^{-2} $ by definition gives us the unit velocity gauge with the unphysical metric induced by $g_{ij}=\CV^{-2}\cg_{ij} $.
The conformal parameter $\t$ becomes the unphysical proper time. Thus if $\t$ is finite at the ideal endpoint of an incomplete conformal geodesic then the singularity can be reached in finite unphysical proper time in the associated unit velocity gauge. This will be a necessary condition for a singularity to be a conformal gauge singularity. 

If $\t$ is not finite, it may be possible to make it finite by a fractional linear transformation. In turn this will be impossible if $\t$ has infinitely many poles accumulating at the singularity. Precisely this behaviour occurs in the Einstein static cylinder, which is the manifold  $\bf{R}\times\bf{S}^3$ with Lorentzian product metric:
$$\tilde{g}=dt^2-dr^2-\sin^2r(d\theta^2+\sin^2\theta d\phi^2).$$
The generators with $(r,\theta,\phi)$ constant are conformal geodesics on which any conformal parameter takes the form $\t=2\tan ((t-t_0)/2)$ for constant $t_0$. The metric $g=t^{-4}\tilde{g}$ has an apparent singularity at $T:=t^{-1}=0$ but every conformal parameter on every generator has infinitely many poles on the approach to it. Thus a conformal rescaling aiming to extend through the singularity will actually push it off to infinite distance. An infinite number of cycles in $\tau$ along a conformal geodesic gives a conformally-invariant notion of \emph{infinitely far way}.

\subsection{Weyl propagation}

Each connection $\Dhat$ defines a parallel transport along a general curve $\c$. We can express the propagation law in a new connection $\D= \Dhat - b$ in terms of $\D$, $b_i$ and $S\tensor{ij}{kl}{}$ using (\ref{Weylsectiontrans}, \ref{tensorsectiontransform}). In the connection $ \D$, we refer to this transport as Weyl propagation or $b$-propagation, where $b$ gives the specific 1-form used for the connection translation. This type of transport can be defined for any connection or 1-form $b$ and any curve $\c$. In the case of a conformal geodesic we always use the 1-form associated to the curve. 

From (\ref{Weylcgdef}), respectively (\ref{Cgequ1}), we see that the velocity of a conformal geodesic $\c$ is by definition Weyl propagated. Let $\{e_\a\}$ be a Weyl propagated frame along  $\c$. The scalar functions $\eta_{\a\b} = \cg(v,v)^{-1}\cg(e_\a,e_\b)$ are conformally invariant and constant along $\c$. Hence angles and length relative to $v^i$ will be preserved by Weyl propagation. 
 We call $\{e_\a\}$ a conformal frame, and call it conformally orthonormal if \mbox{$\eta_{\a\b}=\mathrm{diag}(1,-1,\dots,-1)$}. 
A vector $ e^i$ is Weyl propagated along a conformal geodesic, i.e. it satisfies $\Dhat_v e^i =0$ if and only if in the $\Dhat$-gauge there exist a tractor $E^I$ with components $E_I = (0, \, \frac{1}{\CV}e^i, \, 0) $ satisfying
\be
\label{Etractor}
  \TD_v E^I=\frac{\cg(v,e)}{\cg(v,v)} A^I, \quad \TD^2_vE^I=0 .
\ee
From the definition it follows that $E_I A^I=0 $ and $E_I Z^I=0 $. Clearly $V^I$, the tractor made from $v^i$, satisfies (\ref{Etractor}). The Weyl propagated frame $\{e_\a \}$ with frame metric $\eta_{\a\b} $ gives a tractor tetrad satisfying (\ref{Etractor}) and $\Tg(E_\a, E_\b)=\eta_{\a\b} $. Thus a conformally orthonormal vector basis $\{e_\a \}$, Weyl propagated along the conformal geodesic induces a canonical pseudo orthonormal tractor basis $\{\TE_\TA \}=\{Z, E_\a, A \}$ with frame metric
\be
\label{tractorframemetric}
  \eta_{\TB\TC}=\left( 
\begin{array}{ccc} 0 & 0 & -1\\ 0 & \eta_{\b\c} & 0 \\ -1 & 0 & 0 \end{array} ,
\right)
\ee
where the tractor frame indices $\TB, \TC $ are associated accordingly. Note that the $-1$ entries arise from the minus sign in the projecting part of the acceleration tractor. We can define a dual tractor frame $\Theta^\TA_J := \Tg_{IJ}\TE_\TB^I \eta^{\TA\TB}$ satisfying $\Theta^\TA_J \TE_\TB^J = \delta^\TA_\TB$ and thus expand a tractor $Y^I = Y^\TA \TE^I_\TA$ in the basis $\TE^I_\TA$, where the frame components $Y^\TA:= Y^I \Theta^\TA_I $ are weight-free and conformally invariant. In the $\Dhat$-gauge the splitting of the tractor bundle agrees with the frame components modulo a factor of $-\CV $ or $ \CV^{-1}$. Later on we will use these frame expansions to show the finiteness of tractors arising in our proofs.

\subsection{Jacobi fields and the conformal Jacobi equation}\label{ssJF}

Let $\Gamma $ be a congruence of conformal geodesics transversal to a hypersurface $\Sigma$ with coordinates $\{\s_1, \s_2, \ldots,\s_{n-1} \}$. We can synchronize the congruence by reparametrising so that $\t=0$ on $\Sigma$. This gives us a coordinate system $\{ \s_\ca \} = \{\t, \s_1, \s_2, \ldots,\s_{n-1} \}$ on a patch $V\subset \real^n$ adapted to the congruence in a neighbourhood $ U$ of $\Sigma$. We denote the corresponding local diffeomorphism by $ \psi: V \to U$. The conformal geodesics are given by $\psi(\t,\s^*_{\ca'}) $, where $\s^*_{\ca'}$ are constants. The separation vectors $\eta_{\ca'} = d\psi (\partial / \partial \s_{\ca'}) $ satisfy ${\mathcal{L}}_v \eta_{\ca'} = [v, \eta_{\ca'}] = 0 $ and ${\mathcal{L}}_{\eta_{\ca'}} \eta_{\cb'} = [\eta_{\ca'}, \eta_{\cb'}] = 0$ (recall the primed indices run from 1 to $n-1$).
We choose a scale, say $\co$, and define $Z_\co= \frac{1}{\co} X^I$ and the coordinate tractors associated to the separation vectors $\eta$ and $ \xi$ by $\JT_{(\co)}^I := \TD_\eta \left(Z_\co^I \right)$ and $\JTtwo_{(\co)}^I := \TD_\xi \left(Z_\co^I \right)$ respectively. Then
\be
\label{Jacobitractorswap}
  (\TD_\xi \TD_\eta -\TD_\eta \TD_\xi)\left(Z^I \right) = 0 \, \implies \,
 \TD_\xi \JT_{(\co)}^I = \TD_\eta \JTtwo_{(\co)}^I
\ee
For two different scales $\co$ and $ \ct = \Om \co $ we have $\co \JT_{(\co)} ^I = \ct \JT_{(\ct)} ^I + (\Ups_i \eta^i) X^I$.

Analogously to the Jacobi equation for geodesics, we can derive two equations for a separation vector field $\eta$ from (\ref{Cgequ1}, \ref{Cgequ2})
\bea
\label{confJac1}
  \D_v^2 \eta ^k &=& R\tensor{ij}{k}{l}v^i \eta ^j v^l - \D_\eta \left(b_j S\tensor{il}{jk}{}v^i v^l \right) , \\
\label{confJac2}
  \D_v \D_\eta b_l &=& -b_k R\tensor{ij}{k}{l} v^i \eta ^j + \D_\eta (P_{kl}v^k) + \half \D_\eta \left(b_j b_k S\tensor{il}{jk}{}v^i \right) ,
\eea
These equations are referred to as the conformal Jacobi equations \cite{Friedrich}. 
In the $\Dhat$-gauge, where the congruence 1-form vanishes, the first equation takes on the appearance of the Jacobi equation for geodesics, namely $\Dhat^2_v \eta^k= \Rhat\tensor{ij}{k}{l}v^i \eta ^j v^l $, while the second equation reads $\Dhat_\eta (\Phat_{kl}v^k) =0 $. The separation vectors $\eta^i_{\ca'} = d\psi (\partial / \partial \s_{\ca'}) $ of our congruence are Jacobi fields. For this reason we refer to the associated coordinate tractors $\JT_\ca^I = \TD_{\eta_\ca} Z^I$ as Jacobi tractors. Expressed in the tractor formalism (\ref{confJac1}, \ref{confJac2}) take the form
\bea
\label{tractorconfJac1}
  \TD_v ^2 \JT^K &=& \TCurv(v,\eta)\tensor{}{K}{L}V^L + \TD_\eta A^K , \\
\label{tractorconfJac2}
  \TD_v ^3 \JT^K &=& \TD_v ( \TCurv(v,\eta)\tensor{}{K}{L}V^L) + \TCurv(v,\eta)\tensor{}{K}{L}A^L ,
\eea
where we have rewritten $\TD_v \TD_\eta A^K $ using (\ref{tractorconfJac1}) to obtain (\ref{tractorconfJac2}).

If we are given a coordinate basis $\{ \eta_\ca \}$ with associate coordinate tractors $\JT_\ca $, then $\{A, \JT_\ca, Z \}$ form a tractor basis. It satisfies $\Tg(A,\JT) = \langle b,\eta \rangle$, $\Tg(Z,\JT)=0$ and $\Tg(\JT_\ca,\JT_\cb) = g(\eta_\ca, \eta_\cb) $.

\subsection{Conjugate points}
\label{ssCP}
It is important to know when the coordinate system $\{ \s_\ca \}$ constructed above is well-defined and doesn't develop caustics in the neighbourhood that we want to analyse. The solutions to the conformal Jacobi equation (\ref{confJac1}) form a vector space and any solution is specified by the initial values of $\eta^i$ and $\Dhat_v \eta^i$. Hence there are $2n$ linearly independent Jacobi fields, among which $2(n-2)$ can be specified to start orthogonal to $\c$ at a given point. Two points $p,q \in \c(\t)$ are said to be conjugate if there exists a non-trivial Jacobi field that is parallel to $v$ at both points. A point $p$ is said to be conjugate to $\Sigma$ if the Jacobi field starts tangent to $\Sigma$  and at $p$ is parallel to $v$.

Our coordinate system is well defined as long as the coordinate basis $\{v, \eta_1,\ldots,\eta_n \}$ remains linearly independent. Suppose the spatial coordinates become linearly dependent at $p$. Then there exists a Jacobi field whose spatial part vanishes at $p$ and hence $p$ is conjugate to $\Sigma$. Alternatively if a linear combination of $ \eta_1,\eta_2,\ldots,\eta_n $ is parallel to $v$ at $p$ then the coordinate system breaks down. Thus if we can show that on a given parameter interval the spatial part of a Jacobi field will not vanish then we can conclude that the coordinates are well-defined for that part of the curve and the map $\psi : V \to U$ is a local diffeomorphism. 

To apply these considerations to the study of conformal gauge singularities we suppose that $\c$ is an incomplete time-like conformal geodesic with a finite number of poles in $b$ before the end. We can choose a point $p$ past the last pole and a conformal parameter such that $p=\c(0) $ and $\t$ is finite on the final segment of $\c$. Choose $\Sigma$ to be a smooth surface passing through $p$, orthogonal to $v$ and vanishing second fundamental form at  $p$. Furthermore let $(b_\Sigma)_i$ be a smooth continuation of $b_i(0)$ across $\Sigma$ and $(v_\Sigma)^i$ a vector field orthogonal to $\Sigma$. Now consider the congruence of all conformal geodesics perpendicular to $\Sigma$ with initial data $(v_\Sigma)^i$ and $(b_\Sigma)_i$. We can reparametrise $\t$ so that $\t=0$ on $\Sigma$ and no poles of $b$ appear on the final segment. We define $\{ \s_\ca \} $ and $\psi: V \to U $ as above. We need to analyse whether the coordinate system is well-defined. We shall find in Section 3 that, given appropriate bounds on the tractor curvature along a conformal geodesic, there are no conjugate points within a distance $T$  depending on the bounds. To define the bounds, we first discuss norms.
 
\subsection{Norms}

To define the boundedness of tensor and tractors we will use the Euclidean norms of the frame components in the frame $e_\a$, respectively $\TE_\TB $. The norms are conformally invariant and given by
\bea
\label{Enorm}
  \norm Y \norm &:=& \sqrt{\sum_\TB \Tg(Y,\TE_\TB)^2} = \sqrt{\sum_\TB Y^2_\TB} \, ,\\  
\label{vecnorm}
  \norm y \norm &:=& \sqrt{\sum_\b g(y,e_\b)^2}  = \sqrt{\sum_\b y^2_\b} .
\eea
Since $ y^\b= \eta^{\b\c}  y_\c$, $ Y^\TB= \eta^{\TB\TC}  Y_\TB$, the definition of the norm is independent of the position of the frame indices. The norm is bounded if and only if all the frame components are bounded. Hence whenever we say that a tractor or tensor quantity is bounded, we mean that its norm is bounded, respectively all the frame components are bounded. 
To compress notation, we write 
\bea
\TCurv\tensor{ij}{K}{L} v^i w^j Z^L= [\TCurv(v,w)Z]^K , \non \\
\TCurv_{ijKL} v^i w^j Y^K Z^L = \TCurv(v,w)(Y,Z) , \non \\
(\TD_\xi (\TCurv_{ijKL} v^i w^j)) Y^K Z^L= \TD_\xi (\TCurv(v,w))(Y,Z) \non. 
\eea
We define the following curvature norms
\bea
  \norm  \TCurv \norm  := \sqrt{\sum_{\mu,\nu} \sum_{\TB, \TC} [\TCurv(e_\mu, e_\nu)(\TE_\TB, \TE_\TC)]^2} , \non\\
\label{curvnormfunctions}
  \norm \TCurv^{(q,k)} \norm  := \sqrt{\sum_{\a_1 \ldots \a_k,\mu,\nu} \sum_{\TB, \TC} [(\underbrace{\TD_v \ldots \TD_v}_{q} \TD_{e_{\a_k}} \ldots \TD_{e_{\a_1}} [\TCurv(e_\mu, e_\nu)](\TE_\TB, \TE_\TC)]^2} .
\eea
Thus  $ \norm \TCurv^{(q,k)} \norm $ will correspond to $q$ derivatives along the congruence and $k$ derivatives in arbitrary directions. When $q=0$ we simply omit the zero and write $ \norm \TCurv^{(k)} \norm  $. Here and in the following we will always use $\mu, \,\nu $ to label the two frame vectors contracted onto the two tensor indices of the tractor curvature. This way we are reminded that they are contracted onto $\TCurv$ before the derivatives are applied, while the tractor indices are contracted afterwards.

Let $\eta^i_\ca$ denote another vector basis.  
We can define a Euclidean norm with respect to this frame as well. 
\begin{lemma}
\label{Equivalence lemma}
  Suppose the frame vectors $\eta^i_\ca$ have bounded and non-vanishing norms (\ref{vecnorm}). Then the two Euclidean norms defined by $\{e_\a \}$ and $\{\eta_\ca \}$ are equivalent.
\end{lemma}
\Proofstart This is clear. 
\Proofend

Below, $\{\eta_\ca \}$ will be chosen to be a coordinate frame. In Theorem \ref{conjugate point theorem} we then prove a lower bound and in Corollary \ref{JT0} an upper bound to show that the norm defined by our chosen coordinate frame is equivalent to that defined by the Weyl propagated frame $\{e_\a \}$.

For the remainder of our calculations we fix $\Dhat$ to be the general Weyl derivative associated to the congruence. As before the unphysical metric associated to the unit velocity gauge is defined as $g_{ij}=\CV^{-2}\cg_{ij}$ and the related 1-form as $b$. All unmarked metric expressions, curvature terms etc. refer to the unphysical metric. Thus $b $ denotes the 1-form connecting the Weyl gauge and the unphysical Levi-Civita connection $\D$. We recall that $\Dhat_i g_{jk} = -2b_i g_{jk}$ and $\langle b, v \rangle =0$. Thus $g_{ij}$ is parallelly propagated along the congruence.

\section{Conformal geodesic congruences and the extension theorem}

From the definition in the Introduction, the tractor curvature will be regular at a conformal gauge singularity, and there will be smooth conformal geodesic congruences with Weyl-propagated frames in which the norms of tractor curvature and its derivatives will be bounded. 
We are led to conjecture the converse:
\begin{conjecture}
If the tractor curvature is suitably behaved in a frame induced by a congruence of conformal geodesics ending at a space-time singularity, then there exists a metric $g$ in the conformal class that is well behaved at the singularity and is extendible beyond.  
\end{conjecture}

In trying to prove this, we are motivated by a local extension theorem of R{\'a}cz (p.2459 in \cite{Racz}). Since R\'acz restricts to dimension $n=4$, from now on we shall do the same but there is no reason to suppose that this is necessary. R\'acz's theorem is:
\begin{theorem}
\label{Racz theorem}
Let $\c :(t_1,t_2) \to M$ be an incomplete inextendible time-like geodesic in $(M,g)$. Let $\Wset \subseteq M$ contain a final segment $\c(\bar{t},t_2)$ of $\c$ such that the strong causality condition holds at points of $\c(\bar{t},t_2)$ in the space-time $(\Wset,g\rest _\Wset)$.

\noindent a) Suppose further that there exists $K_0 \in \real^+$ such that the sectional curvature $K(t)$ satisfies $K(t) \ge - K_0$ along $\c$, then there exist:
\begin{enumerate}
\item $t_0 \in (t_1,t_2)$
\item a neighbourhood $\Uset$ of $\c(t_0,t_2)$
\item a map $\phi:\Uset \to \real^4$
\end{enumerate}
such that $\phi$ is an embedding of $\Uset$ into $\real^4$

\noindent b) Suppose further that the components of the Riemann tensor and its covariant derivatives up to order $k+1$ are uniformly bounded 
\footnote{The wording is quoted directly from \cite{Racz}; we shall omit the `uniformly'.}
 in an orthonormal frame field on $U$ 
\footnote{R{\'a}cz constructs a geodesic congruence on $U$ in his proof and parallelly propagates an orthonormal tetrad along it.}.
 Then there exists a nonsingular metric $g^*$ on some neighbourhood $U^*$ of $\overline{\phi[U]}$ such that the map $\phi^* :(U,g\rest_U) \to (U^*,g^*\rest_{U^*})$ is a $C^k$ extension of $(U,g\rest_U)$ into $(U^*,g^*\rest_{U^*})$ i.e. $(M,g)$ is locally $C^k$ extendible.
\end{theorem}
\noindent {\bf Remark}: A set $W$ in a space-time satisfies the strong causality condition if any future-directed causal curve that leaves the set $W$ does not re-enter $W$ at a later time.

\noindent The above theorem is in turn based on the following extension theorem for functions, that combines the results of Whitney \cite{Whitney1,Whitney2} 
\begin{theorem}\label{Whitney theorem}
Let $f$ be a $C^{k+1}$ function on a subset $A \subset \real^n$ satisfying property $\mathcal{{P}}$ {\rm{(defined below)}}, and suppose that the $(k+1)^{th}$ derivatives of $f$ are bounded on $A$, then there exists a function $F$ on $\real^n$ such that
\ben
\item $\partial^m_{x_i} F= \partial^m_{x_i} f $ on $A$, for any multi-index $m =(m_1,\ldots,m_n)$ with $0\leq\Sigma m_i\leq k$;
\item $F$ is analytic on $\real^n \setminus (A \cup \partial A)$.
\een
\end{theorem}
\noindent Whitney's property $\mathcal{{P}} $ is satisfied by the connected set $A$ if there exists a positive  constant $r$ such that for any two points $x, y$ with Euclidean distance $d$ in $\real^n$ there exists an arc in $A$ which connects $x, y$ and has length $d \times r$ or less. 

Our main purpose now is to prove the following related local extension theorem for the conformal structure, following the style of Theorem \ref{Racz theorem}:
\begin{theorem}
\label{Maintheorem}
Let $\c:[0,\t_F) \to M$ be the final segment of an incomplete time-like conformal geodesic in $(M,\tilde{g})$, such that $b$ is bounded in $[0,\t_F) $. Let $W \subset M$ be a neighbourhood of $\c[0,\t_F) $ in which the strong causality condition holds. Let $\{e_\b \}$ be a Weyl propagated orthonormal frame along $\c$ with associated tractor frame $\TE_\TB $. 

\noindent 
 i) Suppose the tractor curvature $\TCurv$ 
has bounded norm $\norm\TCurv\norm$  along $\c$ with respect to $\{e_\b \}$ and $\TE_\TB $. Then there  exists a neighbourhood $U$ of $\c[0,\t_F)$ with $\overline{U} \subset W $ and a diffeomorphism $\psi: V \subset \real^4 \to U $. 

\noindent ii) Suppose further that the tractor curvature norms $ \norm \TCurv^{(1,k+1)} \norm  $ and $ \norm \TCurv^{(k+1)} \norm  $ 
 of equation (\ref{curvnormfunctions}) are bounded on $U$ . Then there exists a general Weyl connection $\Dhat$ associated to $[\tilde{g}]$ and a conformally related nonsingular metric $g$ on some neighbourhood $U^*$ of $\overline{\phi[U]}$ such that the map $\phi^* :(U,g\rest_U) \to (U^*,g^*\rest_{U^*})$ is a $C^k$ extension of $(U,g\rest_U)$ into $(U^*,g^*\rest_{U^*})$.

\noindent iii) The Riemann curvature of $g$ is $C^{k-1} $. 

\noindent Thus the conformal structure $(M,\cg) $ is locally extendible. 
\end{theorem}

For the extension of the metric we will later apply Whitney's extension theorem 
(Theorem \ref{Whitney theorem}) to the components of the tractor metric and their derivatives.  We will start our proof by building a coordinate system around the incomplete time-like conformal geodesic as in Section \ref{ssJF}. As discussed in Section \ref{ssCP}, we need to show that each point on $\c$ has a neighbourhood free of conjugate points on which the map $\psi $ is a diffeomorphism. Let $\chi_{ij}$ denote the second fundamental form of the initial 3-surface $\Sigma$ with respect to the metric $g_{ij}$. 
\begin{theorem}{\textbf{(Conjugate point theorem)}} \label{conjugate point theorem}
\newline Suppose the norm 
 $\norm\TCurv\norm$ is bounded in the Weyl propagated tractor and vector frames along a time-like conformal geodesic $\c(\t)$. Suppose $ \hat{P}$ and $\chi$ denote the norms of $\hat{P}_{ij} $ and $\chi_{ij}$ at $\c(0)$ and that $\TCurv_0 $ is a common bound for $\hat{P},\, 4\chi$ and $\TCurv$. Then there exists a constant $T$ depending on the bound  $\TCurv_0$ such that there are no conjugate points in the interval $[0, T]$ and without loss of generality $\psi : V \to U$ defined in Section \ref{ssCP} is a local diffeomorphism to a neighbourhood $U$ of a final segment of $\c(\t)$.
 
In particular if we assume $\TCurv_0 \ge 2$ then we can set $T= \frac{1}{\TCurv_0} $
\end{theorem}
\Proofstart The idea of the proof is to derive an ODE for $z$ and establish the result via a comparison theorem for $z(\t)$ in the following way.
We derive a differential inequality for $z(\t)$ (\ref{zODE}) and solve the related differential equation in $y_\e(\t)$ with initial data (\ref{yODE}). We show that on an interval $[0,T_\e] $ we have $0<y_\e(\t)<z(\t)$. Thus we conclude that we cannot have conjugate points on $[0,T_\e] $. Since the Schouten tensor is not part of the tractor curvature we will express it as an integral of the Cotton-York tensor, which is bounded by assumption.

As discussed above it is sufficient to show that on an interval $[0, T]$  the vector $\etaperp^i = \eta - g(\eta,v) v^i $ cannot vanish. The frame components of $\eta$ are given by $\eta_0 = \Tg(\JT,V)=g(\eta,v)$ and $\eta_\ap = \Tg(\JT,E_\ap)=g(\eta,e_\ap)$. We write $\etaperp^i= z m^i $, where $z$ denotes the length and $m$ is a space-like unit vector. Let $h_{ij}= -(g_{ij} - v_i v_j) $ denote the positive definite metric orthogonal to the curve, and $h_{\ap\bp} = -h(e_\ap,e_\bp)$ its frame components. For scalars we denote $\frac{d}{d\t} = \Dhat_v = \TD_v $ by $D$ or a dot.

From (\ref{tractorconfJac1}, \ref{tractorconfJac2}) we deduce
$\int^\tau_0 \hat{Y}_{0\b\a}\eta^\b \mbox{d}\sigma = \hat{P}(\eta,e_\a)(\tau) - \hat{P}(\eta,e_\a)(0)$ and
\bea
\label{etaODE}
  D^2\eta_\ap =  
\TCurv(v, e_\bp)( E_\ap,V)\eta^\bp + \int^\t_0 \TCurv(v , e_\bp)( E_\ap,A)\eta^\bp ds + \hat{P}(\eta, e_\ap)_*, \\
  D^3 \eta_0 =  -\widehat{\CY}(v,e_\bp,v)\eta^\bp. \phantom{\TCurv(v, e_\bp)( E_\ap,V)\eta^\bp\,  \int^\t_0 \TCurv(v , e_\bp)( E_\ap,A)\eta^\bp d}
  \eea
We define the quantities
\bea
\label{TCurvmatrices}
  A_{\ap\bp} &=& \TCurv(v, e_\bp)( E_\ap,V)=C_{ijkl}v^i e^j_\bp e^k_\ap v^l ,\non \\
  B_{\ap\bp} &=& \TCurv(v, e_\bp)(E_\ap,A)=\widehat{\CY}(v,e_\bp , e_\ap) =\widehat{\CY}_{ijk}v^i e^j_\bp e^k_\ap , \non \\
  C_{\bp} &=&  \TCurv(v, e_\bp)(V,A)=\widehat{\CY}(v,e_\bp , v) =\widehat{\CY}_{ijk}v^i e^j_\bp v^k .
\eea
We note that $A_{\ap\bp} = - E_{\ap\bp}$ is (minus) the electric Weyl tensor in the unit gauge. $A_{\ap\bp}, B_{\ap\bp}, C_\bp$ are tractor curvature components and hence
 $\norm A \norm$, $\norm B \norm$, $\norm C \norm \le  \norm \TCurv \norm \le \TCurv_0$. For our calculations it is convenient to use the more specific bounds $\norm A \norm\le K$, $\norm B \norm \le L$.

From $z^2= h_{\ap\bp}\eta^\ap \eta^\bp $ we deduce
\bea
  z\dot{z} &=& h(D \etaperp, \etaperp) = \chi(\etaperp, \etaperp) , \non \\
  \ddot{z} &=& \frac{h(D^2 \etaperp, \etaperp)}{z}+\frac{h(D\etaperp, D\etaperp)h(\etaperp, \etaperp) - (h(D \etaperp, \etaperp))^2}{z^3} \ge \frac{h(D^2 \etaperp, \etaperp)}{z} \non
\eea
having used the Cauchy-Schwarz inequality on the second term. Now by substituting (\ref{etaODE}) and then applying (\ref{TCurvmatrices}) with their bounds we get
\bea
   \ddot{z}(\t)  &\ge &  - z(\t) (A_{\ap\bp}m^\ap m^\bp)(\t) - \int^\t_0 m^\ap(\t)(B_{\ap\bp}m^\bp)(s)z(s) ds \non \\
   & &+ \hat{P}_{\ap\bp}m^\ap(0)m^\bp(\t) \non \\
\label{zODE}
                      &\ge & -K z(\t) - L \int^\t_0 z(s)ds - \hat{P}.
\eea
Since the conformal Jacobi equation (\ref{confJac1}) is linear it is sufficient to consider initial data of the form
$$\eta^i_*= e^i_\a, \quad z_*=1 ,\quad X:=\dot{z}_*=\chi_{\a\a}(p).$$
Hence there exists an interval on which $z$ is positive. Note that $-\chi \le X \le \chi$.
It follows from (\ref{zODE}) that $\ddot{z}(0) \ge -(K+\hat{P}) $ with equality when $\chi_{ij} $ vanishes at the inital point $p$. The latter can be achieved using the exponential map associated to $g_{ij}$ at $p$.

Let $y_\e(\t)$ be a solution for the initial value problem.
\bea
\label{yODE}
  \ddot{y}_\e(\t) = -(K + \e^2) y_\e(\t) - L \int^\t_0 y_\e(s)ds - \hat{P}, \non \\
  y_{\e*}=1,\,\,\, \dot{y}_{\e*}= X,\,\,\, \ddot{y}_{\e*}=  -(K+\hat{P}+\e^2).
\eea

We will drop the $\e$-subscript for a moment and assume $\e>0$ is implied. It follows that $y>0$ on some interval $I=(0,T]$. To show that $z>0 \phm \forall \t\in I$, we compare $z$ to $y$ by analysing the ratio $R=\frac{z}{y}$ on $I$. Define $W :=\dot{z}y - \dot{y}z = y^2 \dot{R}$. We have $R_*=1$ and $W_*=0$. Combining (\ref{zODE}) and (\ref{yODE}) one gets for $\dot{W}(\t) = y\ddot{z} - z\ddot{y}$ that
\be
\label{WODE}
  \dot{W}(\t)  \ge \e^2 y(\t)z(\t) + L \int^\tau_0 y(\tau)y(\sigma)  [R(\tau)-R(\sigma)] d\sigma + \hat{P}y(\tau)[R(\tau)-1] .
\ee
Observing that $\dot{W}_* \ge \e^2$ it follows that $W$ is non-negative on an interval $J:=[0,c_1] \subseteq I$. Thus $\dot{R}>0$ on $J$. This implies that $R$ is increasing and by (\ref{WODE}) $\dot{W}(c_1)>0$. So $W$ is positive beyond $c_1$. Clearly, as long as $R$ is defined $W$ will have no maximum and hence no zero. Therefore we conclude $W>0$ and $R\ge 1$ on $I$. This implies that for each $\e>0$, $z$ is positive at least until the first positive zero of $y_\e(\t)$, at $T_\e$ say. In the limit $\e \to 0$, $y_\e(\t) \to y_0(\t)$ and $T_\e \to T_0>0$. Take $T=T_0$, then $y_0$ and hence $z$ will not vanish on $[0,T]$, which concludes the proof, once we have a lower bound on $T$ in terms of the curvature bound.

To obtain a lower bound on $T$ in terms of the positive bounds $\hat{P},\, \chi, \, K, \,L$ we proceed as follows.
From (\ref{yODE}) with $\e=0 $ we get
\bea
\label{ydotintegral}
  \dot{y}(\t) = X - \hat{P}\t - K  \int^\t_0 y(\s)d\s -  L  \int^\t_0 (\t-\s) y(\s)d\s ,\\
\label{yODEsolution}
  y(\t) = 1 + X\t - \half \hat{P}\t^2 - K  \int^\t_0 (\t-\s)y(\s)d\s - \half L  \int^\t_0 (\t-\s)^2 y(\s)d\s .
\eea
It follows that $\dot{y} \le X $ and $y \le 1 + X\t \le 1+ \chi \t $, since $-\chi \le X \le \chi$. Substituting into (\ref{yODEsolution}) and integrating gives
\be
y(\tau) \ge Y(\tau)\equiv 1 -  \chi \tau - \frac{1}{2} (\hat{P}+K)\tau^2 - \frac{1}{6}(\chi K+L)\tau^3 - \frac{L\chi }{24}\tau^4
\ee
$Y(\t)$ must have a unique positive root. Setting 
\be 
q = max( 4\chi, (\hat{P}+K)^{\half} , (\chi K+L)^{\frac{1}{3}} ,(\chi L)^{\frac{1}{4}}) 
\ee
we get $Y(\frac{1}{q})\geq \frac{1}{24} $. Thus under our additional assumption of $\TCurv_0 \ge 2 $ we have $\frac{1}{\TCurv_0} \le \frac{1}{q}  $ and hence $z(\TCurv_0^{-1}) \ge y(\TCurv_0^{-1})\ge Y(\TCurv_0^{-1}) >0$. 

\noindent Hence in this case $\frac{1}{\TCurv_0} $ gives the desired lower bound for $T$.
\Proofend

\textbf{Remark:} For $\chi=0 $ the function $Y(\t)$ is a cubic and thus solving it for the positive root it might be possible to obtain an improved estimate.

This proves part (i) of Theorem \ref{Maintheorem}. For part (ii) we need Whitney's Property $\mathcal{{P}} $, to hold for (a refinement of) $U$. 
R{\'a}cz constructs a neighbourhood ${\mathcal{O}} \subset \psi^{-1}[W \cap U] \subset V$, satisfying Whitney's property $\mathcal{{P}} $, for the final part of the incomplete curve $\c$ in Proposition 3.2.6 of \cite{Racz}, where $W$ is a neighbourhood of $\c$ in which the strong causality condition holds.  Strong causality is an invariant property of the conformal class $[g]$, so we identify $W$ using the physical metric. By the above theorem any point $q$ in the final segment must have an open neighbourhood ${\mathcal{O}}_q \subset W $ on which $\psi $ is a diffeomorphism. For the construction of ${\mathcal{O}} $ we uses the fact that each ${\mathcal{O}}_q $ must contain a causally convex neighbourhood $w_q \subset {\mathcal{O}}_q $. For each $q$ one then fits an open 3-ball $ B_q(\r)$ of radius $\r$ centred at $\psi^{-1}(q) $ lying in the surface of constant $\t$ into $\psi^{-1}(w_q) $. The radii are chosen such that $\r$ is nowhere vanishing and decreasing with $\t$. One defines the interior of the union of the 3-balls over $[0, \t_F] $ as the set ${\mathcal{O}}$ given above. It follows that $\psi\rest_{\mathcal{O}}$ must be injective. Furthermore this neighbourhood ${\mathcal{O}}$ satisfies Whitney's property $\mathcal{{P}} $ (Lemma 3.3.2 \cite{Racz}). This is the local neighbourhood on which we want to work. 
 Thus we have shown
\begin{lemma}
\label{coordinate system}
Given an incomplete time-like conformal geodesic $\c$ with bounded norm 
$\norm\TCurv\norm$ of the tractor curvature,  we can construct a coordinate system around a final segment of $\c$ such that the property $\mathcal{P}$ is satisfied in a neighbourhood of the final segment.
\end{lemma}
The rest of the proof of Theorem \ref{Maintheorem} needs an intricate series of inductions, which we turn to next.

\section{ Extending the conformal metric}
\label{Proof}

To apply Whitney's Theorem we need to show that the derivatives up to order $k$ of the components of the tractor metric are bounded  and can be extended onto the boundary of our chosen set $\mathcal{O} $. This will be guaranteed by having the derivatives up to order $k+1$ bounded on $\mathcal{O} $, as was shown in Lemma 3.3.1 of \cite{Racz}. 

We observe the following relationship between the tractor metric and the metric $g_{ij}= \CV^{-2}\cg_{ij}$ induced by the congruence.
\[
  \Tg(\JT_\ca, \JT_\cb) =\eta_{\a\b} \JT^\a_\ca \JT^\b_\cb = g(\eta_\ca, \eta_\cb) \]
so that
 \bea\label{metricderivatives}
 \TD_{\xi_j} \ldots \TD_{\xi_1} \Tg(\JT_\ca, \JT_\cb) &=& \TD_{\xi_j} \ldots \TD_{\xi_1} [\eta_{\a\b}\JT^\a_\ca \JT^\b_\cb] \non\\
&=& \Dhat_{\xi_j} \ldots \Dhat_{\xi_1}g(\eta_\ca, \eta_\cb) ,
\eea
where we take derivatives along the coordinate fields $\xi_i \in \{ \eta_\ca \}$. To prove the boundedness of the derivatives of the metric components (\ref{metricderivatives}), we expand these expressions. Since $\Tg$ and $\eta_{\TB\TC} $ are preserved by the tractor connection it is enough to show that the Jacobi fields represented by $\JT^I_\cb $, respectively their components $\JT^\TB_\cb$ with respect to the tractor basis, have bounded derivatives up to order $k+1$ along the coordinate fields $\xi_i $. For the extension it is thus sufficient to prove that $ \TD_{\xi_{k+1}} \ldots \TD_{\xi_1} \JT^\TB_\cb$ are 
 bounded.

\subsection{Notation and outline of the proof}
\label{outline of main proof}

We now prove part (ii) of Theorem \ref{Maintheorem}. Since it requires a long induction process, we start by setting up simplifying notation and giving an outline of the proof. 
Formulas will often be written in a semi-schematic way. The coordinate frame will always be denoted $\eta$ with corresponding tractor $\JT$. Any coordinate frame vector in a derivative will be denoted by $\xi$. $\xi$ will be used both as a single and as a multi-index. The latter may be broken up into parts, each denoted $\xi$ as well. Hence two $\xi$'s in an expression can be of different nature and value. Where it matters we will be more specific, but in general we will assume that the exact form follows from the equations and the context. 

\noindent For higher order derivatives we will write
\bea
  Y^I_{(j)} = \TD^j_\xi \JT^I= \TD_{\xi_j}\ldots \TD_{\xi_1}\JT^I , \qquad Y^I_{(0)}=\JT^I ; \non \\
\label{yreplace}
  y^\TB_{(j)} = \TD^j_\xi (\JT^\TB) = \TD_{\xi_j}\ldots \TD_{\xi_1}(\JT^\TB), \qquad y^\TB_{(0)}=\JT^\TB ;
\eea
and it is implied that $Y^I_{(j)} = \TD_{\xi_j} Y^I_{(j-1)}$ and $y^\TB_{(j)} = \TD_{\xi_j} y^\TB_{(j-1)}$. When $q$ time derivatives are applied to a general $j^{th}$ derivative we refer to it as a $(q,j)$-derivative, e.g. $\TD^q_v \TD^j_\xi \JT$.

Our overall aim is to show that  $\TD^q_v \TD^l_\xi\JT^\TB$ are bounded for $q \le 3, l \le k+1$. 
We start with the derivation of a generalised version of the conformal Jacobi equations (\ref{tractorconfJac2}) to derive an upper bound for $\TD^q_v \TD^l_\xi\JT^\TB$. 
This is done in Theorem \ref{Upperboundtheorem}, the central part in the proof of the extendibility. It is stated and proven first with the necessary conditions shown to hold afterwards. Their proof by induction has been split into several lemmas.
 The $j^{th}$ inductive step assumes bounded $\norm\TCurv^{(1,j)}\norm $, $\norm\TCurv^{(j)}\norm$ and the general structure is outlined below.
\bitem
\item Lemma \ref{DxiEk} [1,$j$] deduces the boundedness of $\TD^3_v\TD^{j}_\xi \TE_\TB$
\item Lemma \ref{Qk} [2,$j$] shows that the source terms $Q_{(j)} $ and $\bar{Q}_{(j)}$ in (\ref{Jacobialpha} - \ref{JacobiZ}) are bounded
\item Lemma \ref{JTk} [3,$j$] combines the generalised conformal Jacobi equation for tractors and Theorem \ref{Upperboundtheorem} to show that $\JT^\TB$ has bounded $(q,j)$ derivatives for $q \le 3$.
\eitem

In preparation, we derive a few formulas and two lemmas that will be used in the proofs and make a few general observations. Most intermediate proofs will combine an expansion of the frame components with the integration lemmas \ref{integrateDy} and \ref{integrateDY} below. In general, we isolate an unknown term and show that all the remaining terms in the expansion are bounded, by applying bounds from lower levels of induction and differentiation. 

To permute a time derivative with $j$ coordinate derivatives acting on a tractor $ \TE^I$, respectively a vector $e^i_\a $, we use
\bea
\label{curvswapE}
  \TD_v \TD^j_\xi \TE^I = \sum^{j-1}_{l=0} \TD^l_\xi [\TCurv(v,\xi_{j-l})\TD^{j-(l+1)}_\xi \TE]^I + \TD^j_\xi \TD_v \TE^I , \\
\label{curvswape}
  \Dhat_v \Dhat^j_\xi e^i = \sum^{j-1}_{l=0} \Dhat^l_\xi [\Rhat(v,\xi_{j-l})\Dhat^{j-(l+1)}_\xi e_\a]^i + \Dhat^j_\xi \Dhat_v e^i  .
\eea

One of our assumptions in Theorem \ref{Maintheorem} is that the norms $\norm\TCurv^{(1,k+1)}\norm$ and hence the components of $\TD_v \TD^{k+1}_{e_\a} \TCurv\tensor{\mu\nu}{K}{L}$ are bounded with respect to the tractor frame $\{\TE_\TA \}$. The derivatives are taken along the conformally orthonormal frame $\{ e_\a \}$. Our calculations on the other hand will mainly contain derivatives and frame components with respect to the coordinate frame  $\{\eta_\ca \}=\{\xi_\ca \} $. To be able to deduce the boundedness of these derivatives from these norms we need to replace all $\TD_\xi$'s by $\TD_{e_\a}$'s. We use $\eta^i=\eta^\a e^i_\a = \Tg(\JT, E_\a)\eta^{\a\b}e^i_\b $ to rewrite the derivatives
\bea
\label{TCurvxitoe}
  \TD^j_\xi\TCurv\tensor{\mu\nu}{K}{L}
  &=& \sum_{\b_1}\TD^{j-1}_\xi\left(\xi^{\b_1}\TD_{e_{\b_1}}\TCurv\tensor{\mu\nu}{K}{L} \right)  \non \\
  &=& \sum_{\b_1,\ldots,\b_j} \xi^{\b_j} \TD_{e_{\b_j}}\left( \ldots \left(\xi^{\b_2}\TD_{e_{\b_2}}(\xi^{\b_1}\TD_{e_{\b_1}} \TCurv\tensor{\mu\nu}{K}{L}) \right) \right) .
\eea
Expanding the above formula, we get $\TD^l_\xi \xi^{\b_m}$ with $l \le j-1$, which further give rise to terms involving $\TD^l_\xi \xi^I$ and $\TD^l_\xi E_\a^I$ . By applying $\TD_v$ to (\ref{TCurvxitoe}), one sees that the boundedness of $\TD_v \TD^k_\xi \TCurv\tensor{\mu\nu}{K}{L}$ follows from bounded $\TD_v \TD^l_\xi \xi^I$ and $\TD_v \TD^l_\xi E_\a^I$ for $l \le j-1$.

\begin{lemma}
\label{integrateDy}
  Suppose $y^i$ has the norm $\norm \Dhat_v y^i \norm$ of its derivative bounded on an interval $[\t_0 , \t_0 + T]$ of $\c(\t)$ and its initial value $\norm y^i \norm \rest_{\t_0}$ also bounded. Then $\norm y^i \norm$ is bounded on $[\t_0 , \t_0 + T]$.
\end{lemma}
\Proofstart Denote the bounds by $F_1$ and $F_0$ respectively. Since $g_{ij}$ is preserved along the curves and the frame $e_\a$, $\a \in \{0,1,2,3 \}$ is $b$-propagated we get $\frac{d}{d\t}g(y,e_\a) = g(\Dhat_v y, e_\a) \le F_1$. Integrating over the interval $[\t_0, \t_0 + T] $ we get 
\be
  g(y,e_\a)\rest_\t=g(y,e_\a)\rest_{\t_0} + \int_{\t_0}^{\t} (g(\Dhat_v y, e_\a))ds \le F_0 + F_1 T . \non
\ee
Now $g(\Dhat_v y,e_\a) = (\Dhat_v y)_\a$ are bounded, which implies that $y^i$ has a bounded norm.
\Proofend
\begin{lemma}
\label{integrateDY}
  Suppose $Y^I$ has the norm $\norm \TD_v Y^I \norm$ of its derivative bounded on an interval $[\t_0 , \t_0 + T]$ on $\c(\t)$ and its initial value  $\norm Y^I \norm_{\t_0}$ also bounded. Then $\norm Y^I \norm$ is bounded on $[\t_0 , \t_0 + T]$.
\end{lemma}
\Proofstart Again denote the bounds by $F_1$ and $F_0$. We have $\frac{d}{d\t}\Tg(Y,\TE_\TB) = \Tg(\TD_v Y, \TE_\TB) + \Tg(Y, \TD_v \TE_\TB) $. For $\TE_\TB=A$ or $\TE_\TB=E_\b$ we get
\be
  \Tg(Y,\TE_\TB)\rest_\t=\Tg(Y,\TE_\TB)\rest_{\t_0} + \int_{\t_0}^T (\Tg(\TD Y, \TE_\TB))ds \le F_0 + F_1 T . \non
\ee
Using this bound we can get a bound for $\TE_\TB = V$ and from that for $\TE_\TB = Z $
\bea
   \Tg(Y,V)\rest_\t &\le& F_0 (1+T) + F_1\left(T + \half T^2\right) , \non \\
   \Tg(Y,Z)\rest_\t &\le& F_0 (1+T+\half T^2) + F_1\left(T + \half T^2 + \frac{1}{6}T^3\right) . \non
\eea
Hence $\norm Y^I \norm _\t \le \sqrt{6}(F+F_1)e^T$ on $[\t_0 , \t_0 + T]$. \Proofend

By repeated application of the lemma \ref{integrateDY} we can see that given bounded $\norm\TCurv^{(1,j)}\norm $, we have bounded $\norm\TCurv^{(l)}\norm$ and $\norm\TCurv^{(q,l)}\norm$ for $0 \le l+q \le j $, as we can set $e_\a=v$. It follows from (\ref{TCurvxitoe}) that $\TD_v \TD^j_\xi \TCurv $ and $\TD^q_v \TD^l_\xi \TCurv $ for $0 \le l+q \le j $ are bounded, as long as we have bounded $(1, j-1)$-derivatives for the Jacobi tractor $\JT^I_\cb$ and its components $\JT^\TB_\cb$.

Before we start the actual proof, we expand the tractor  $Y^I_{(j)}=\TD^j_\xi \JT^I$ in terms of the functions $y^\TB_{(l)}=\TD^l_\xi (\JT^\TB)$ and $\TD^m_\xi \TE^I_\TB $. We define 
\bea
  T^I_{(j)} := (\TD^j_\xi \JT^\TB)\TE^I_\TB = y^\TB_{(j)} \TE^I_\TB,  \non \\
  \label{defTTbar}
  \bar{T}_{(j)} :=Y_{(j)}-T_{(j)}, \quad \bar{Q}^I _{(j)}:=\TD^3_v \bar{T}^I_{(j)}
\eea
It follows that $\TD^3_v Y^I_{(j)}=\TD^3_v T^I_{(j)}+\bar{Q}^I _{(j)}$. We can see by expansion that $\bar{T}_{(j)} $ contains $y^\TB_{(l)}, \, l<j,$ and $\TD^m \TE^I_\TB, \, m \le j $ and $\bar{Q} _{(j)}$ contains their time derivatives up to order $q=3 $.

Using (\ref{tractorconnection}) in the $\Dhat$-gauge, the components of the tractor $Y_{(j)}=\TD^j_\xi \JT$ contain derivatives of $\eta$ and $\xi$ and contractions with $b$, $\Phat$ and their derivatives. The same holds for $\TD^j_\xi \TE_\TB$. The highest order terms are $\Dhat^j_\xi \eta$ and  $\Dhat^j_\xi \langle b, \eta \rangle $. This will be used later to prove the boundedness of these terms.

\subsection[The generalised Jacobi equation and the upper bound theorem]{The generalised Jacobi equation and \\ the upper bound theorem}

We begin with a generalised conformal Jacobi equation for the $Y_{(j)}$:
\begin{proposition}
  Let  $Y_{(j)} = \TD^j_\xi \JT$ denote the $j^{th}$ derivatives of the Jacobi tractor $\JT$ along the coordinate fields and let $y^\b_{(j)} = \Dhat^j_\xi (\eta^\b)$ be the derivatives of the frame components of the Jacobi field. Then $Y_{(j)}$ satisfies the following generalised conformal Jacobi equation for tractors
\be
\label{generalconfJac} 
  \TD_v ^3 Y_{(j)}^K = \TD_v ( \TCurv(v,e_\b)\tensor{}{K}{L}V^L y^\b_{(j)}) + \TCurv(v,e_\b)\tensor{}{K}{L}A^L y^\b_{(j)} + Q_{(j)}^K ,
\ee
where $Q_{(j)}^K$ is defined recursively by
\be
  Q_{(0)}^K=0 ,
\ee
\bea
\label{Qkformula}
 Q_{(j)}^K &=& \TD^2_v (\TCurv (v, \xi )\tensor{}{K}{L} Y_{(j-1)}^L )  + \TD_v (\TCurv (v, \xi )\tensor{}{K}{L}\TD_v Y_{(j-1)}^L ) + \TCurv(v, \xi)\tensor{}{K}{L}\TD_v^2 Y_{(j-1)}^L  \non \\
& & + \TCurv(\xi ,v)\tensor{}{K}{M}\TCurv(v,e_\b)\tensor{}{M}{L}V^L y^\b_{(j-1)} + \TD_\xi (\TCurv(v,e_\b)\tensor{}{K}{L}A^L)y^\b_{(j-1)} \non \\
& & + \TD_\xi (\TCurv(v,e_\b)\tensor{}{K}{L}V^L) \TD_v y^\b_{(j-1)} + \TD_v \TD_\xi (\TCurv(v,e_\b)\tensor{}{K}{L}V^L) y^\b_{(j-1)}  \non \\ 
& & + \TD_\xi Q_{(j-1)}^K .
\eea
\end{proposition}

\Proofstart The proof is inductive, by applying the conformal Jacobi equation and swapping derivatives using curvature terms. Equation (\ref{tractorconfJac2}) gives the case $n=0$.
To deduce $n=j$ from $n=j-1$ we write 
\begin{eqnarray*}
 \TD_v^3 \TD_\xi Y_{(j-1)}^K &=& \TD^2_v (\TCurv (v, \xi )\tensor{}{K}{L} Y_{(j-1)}^L )  + \TD_v (\TCurv (v, \xi )\tensor{}{K}{L}\TD_v Y_{(j-1)}^L )\\
&& + \TCurv(v, \xi)\tensor{}{K}{L}\TD_v^2 Y_{(j-1)}^L  + \TD_\xi \TD_v^3 Y_{(j-1)}^K . \non 
\end{eqnarray*}
This leads to the first 3 terms in (\ref{Qkformula}). We expand the fourth term with the use of (\ref{generalconfJac}) for $j-1$ and swap the derivatives on the first term
\begin{eqnarray*}
 \TD_\xi \TD_v^3 Y_{(j-1)}^K &=& \TCurv(\xi , v)\tensor{}{K}{L}(\TCurv(v,e_\b)\tensor{}{L}{M}V^M y_{(j-1)}^\b) + \TD_v \TD_\xi (\TCurv(v,e_\b)\tensor{}{K}{L}V^L y_{(j-1)}^\b) \non \\
& & + \TD_\xi(\TCurv(v,e_\b)\tensor{}{K}{L}A^L y_{(j-1)}^\b) + \TD_\xi Q_{(j-1)} .
\end{eqnarray*}
Expanding the terms and isolating the ones containing $y^\b_{(j)} = \Dhat_\xi y_{(j-1)}^\b $ we obtain the result.
\Proofend

Next we prove that the derivatives of the metric components (\ref{metricderivatives}) have an upper bound that can be derived from the boundedness of tractor curvature. In Theorem \ref{conjugate point theorem} we proved the existence of a lower bound for the norms of the coordinate frame $\{\eta_A\}$. Now we use similar steps to derive the upper bound for $y^\TB_{(j)} = \TD^j_\xi (\JT^\TB)$. We prove the general case $j$  assuming certain \textit{a priori} bounds. Their existence will be proven by induction later on.
\begin{theorem}{\textbf{Upper bound theorem}\phm}
\label{Upperboundtheorem}
Suppose the following hold: 
\newline i) 
 the norm  $\norm\TCurv\norm$ is bounded, say $ \norm\TCurv\norm \le \TCurv_0$;
\newline ii) there exists a bound $\half q_0 $ for the norms of $Q_{(j)}$, $\bar{Q}_{(j)}$.
\newline Then $y^\TB_{(j)} = \TD^j_\xi (\JT^\TB)$ and its three time derivatives are bounded.
\end{theorem}

\Proofstart  Using the split $\TD^3_v Y^I_{(j)} = \TD^3_v T^I _{(j)}+ \bar{Q}^I_{(j)}$ according to (\ref{defTTbar}) gives differential equations for the frame components of $T^I_{(j)}$ from (\ref{generalconfJac}). Denote $\TD_v$ acting on scalars by $D$ and recall that the tractor frame metric is constant, then (suppressing the subscript $(j)$ to simplify the formulae)

\bea
\label{Jacobialpha}
  D^3 y_\ap =& D(A_{\ap\bp}y^\bp) + B_{\ap\bp}y^\bp &+ \phm Q_\ap - \bar{Q}_\ap ,\\
\label{JacobiA}
  D^3 y_\TAA =& -D(C_{\bp}y^\bp) \phantom{+ B_{\ap\bp}y^\bp} &+ \phm Q_\TAA - \bar{Q}_\TAA ,\\
\label{Jacobi0}
  D^3 y_0 =& 2 C_\bp y^\bp  + 3 D^2 y_\TAA \phm  &+\phm  Q_0 - \bar{Q}_0 , \\
\label{JacobiZ}
  D^3 y_\TZ =& 3 D^2 y_0+ 3 D y_\TAA \phm &+\phm  Q_\TZ - \bar{Q}_\TZ  ,
\eea
where $A_{\ap\bp}$, $B_{\ap\bp}$ and $C_\bp $ as before in (\ref{TCurvmatrices}).
We can see that (\ref{Jacobialpha} - \ref{JacobiZ}) form a hierarchy of differential equations, which can be solved in order. We will derive a comparison theorem for $y_\bp$ in (\ref{Jacobialpha}). Once we have obtained bounds for $y_\bp$, we can substitute them into the other equations, where it is sufficient to show that the right hand side is bounded. 

We define $y^\ap e^i_\ap=y^i_\perp= z n^i$, where $n^i $ is a space-like unit vector. Hence, $z$ is the length of $y_\perp$ and its norm. 
We integrate (\ref{Jacobialpha}) three times and take the norm of both sides to get
\bea
&  &\norm y_\ap(\t)-y_\ap(0)-\t\dot{y}_\ap(0)-  \half \t^2 [\ddot{y}_\ap(0)-(A_{\ap\bp}y^\bp)(0)] \norm \non \\
&=& \norm \int^\t_0 \int^\s_0 (A_{\ap\bp}y^\b)(\s') +\left( \int^{\s'}_0 (B_{\ap\bp}y^\bp + q_\ap)(\s'') d\s'' \right) d\s' d\s \norm \non \\
&\le & \int^\t_0 \int^\s_0 \norm (A_{\ap\bp}y^\bp)(\s') \norm +\left( \int^{\s'}_0 \norm (B_{\ap\bp}y^\bp)(\s'')\norm + \norm q_\ap(\s'')\norm d\s'' \right) d\s' d\s \norm \non \\
\label{yJacobi}
&\le & \int^\t_0 \int^\s_0 a(\s')z(\s') +\left( \int^{\s'}_0 b(\s'')z(\s'')+ g(\s'') d\s'' \right) d\s' d\s ,
\eea
where $q=q_{(j)}=Q-\bar{Q} $ and $a(\t),b(\t),g(\t)$ are bounds for $\norm A \norm, \norm B \norm, \norm q_{(j)} \norm$. We know that $\norm A \norm, \norm B \norm, \norm C \norm$ are 
bounded by $\TCurv_0 $, and $\norm q \norm \le q_0$ follows from our assumptions.

The idea is to use the equation $\frac{d}{d\t}(\ddot{y}-a y ) = by + g$, with $a(\t)>0 $ and $b(\t)>0$ on $[0,T] $ to derive a comparison theorem for $z$. The differential equation in $y$ implies analogously to the above that
\bea
&y(\t)-y(0)-\t\dot{y}(0)-\half \t^2 [\ddot{y}(0)-a(0)y(0)] \non \\
\label{ycomparison}
=& \int^\t_0 \int^\s_0 a(\s')y(\s') +\left( \int^{\s'}_0 b(\s'')y(\s'')+ g(\s'') d\s'' \right) d\s' d\s .
\eea
Subtracting  (\ref{ycomparison}) from (\ref{yJacobi}) we get
\bea
& & \norm y_\ap(\t)-y_\ap(0)-\t\dot{y}_\ap(0)-\half \t^2 [\ddot{y}_\ap(0)-(A_{\ap\bp}y^\bp)(0)] \norm \non \\
&-& \left( y(\t)-y(0)-\t\dot{y}(0)-\half \t^2 [\ddot{y}(0)-a(0)y(0)] \right) \non \\
\label{ymaster}
&\le& \int^\t_0 \int^\s_0 a(\s')[z-y](\s') +\left( \int^{\s'}_0 b(\s'')[z-y](\s'') d\s'' \right) d\s' d\s .
\eea
Now suppose that $y_\e(\t)$ $(\e>0)$ be a solution (\ref{ymaster}) with initial data
\be
\label{yinitdata}
  y_\e(0)=z(0) + \e, \quad \dot{y}_\e(0) = \norm  \dot{y}_{\ap}\norm_0, \quad \ddot{y}_\e(0)-a(0)y_\e(0)= \norm \ddot{y}_\ap - A_{\ap\bp}y^\bp \norm_0 .
\ee
We define $\t_{max}:=sup_{\t \in [0,T]} \{\t : z(\s) \le y_\e(\s)  \forall \s \in [0,\t]\} $. Then the right hand side of (\ref{ymaster}) is negative on $[0,\t_{max}]$. Using $\norm a-b-c-d \norm \ge \norm a \norm - \norm b \norm - \norm c \norm - \norm d \norm$ and the initial data we deduce that 
\be
  z(\t) + \e \le y_\e(\t)  \quad \forall \t \in [0,\t_{max}] . \non
\ee
Hence by continuity $z \le y_\e $ for some $\t \in [\t_{max}, T]$, which leads to a contradiction unless $\t_{max}=T$. Thus for any $\e>0$, $z$ is bounded by $y_\e$ on $[0,T]$. Now taking the limit $\e \to 0$ we see that there exists a solution $y$ of (\ref{ycomparison}) with initial data (\ref{yinitdata}) and $\e=0$ and $z(\t) \le y(\t) \phm \forall \t \in [0, T]$

We established earlier that we can set $a(\t)=\TCurv_0 = b(\t)$ and $g(\t) = q_0$. Thus $z$ is bounded by a solution of $\frac{d}{dt}(\ddot{y}-\TCurv_0  y(t)) - \TCurv_0 y(t) = q_0 $. Since the coefficients are constants we can deduce that the auxiliary equation is a polynomial of order 3 with 3 complex roots $\l_i$. Thus all solutions are linear combinations of $e^{\l_i \t}$ and hence must be bounded on a finite interval. It follows that the norm $z$ of $y_\perp$ is bounded on $[0,T]$ and so are its frame components $y_\ap $. From (\ref{Jacobialpha}) and (\ref{yJacobi}) we deduce that $Dy_\a, D^2 y_\a, D^3 y_\a$ are bounded as well. 

Then using comparison theorems like the above for $y_A, y_0$ and $y_Z$ we deduce that each one and its first three time derivatives must be bounded on $[0,T]$. Putting back the subscript $(j)$, we can conclude that $\TD^q_v y^\TB_{(j)}$ are bounded for $q\le3$. \Proofend

This has proved the boundedness of $\TD^j_\xi \JT^\TB$ for any $j$, as long as sufficient differentiability of the curvature is guaranteed. The case $j=0$ is an immediate corollary. 
\begin{corollary}{\textbf{[3,0]}\phm}
\label{JT0}
  Given $ \norm\TCurv\norm \le \TCurv_0$, 
 we deduce that $\JT$, $\eta$, $\TD_v^q \JT$ and $\Dhat^q_v \eta$ are bounded for $q \le 3$.
\end{corollary}
\Proofstart For $j=0$ we observe that $Q_{(0)}=0$ by definition and since $T^I = \JT^I$ and $\bar{T}^I=0  $ we have $\bar{Q}_{(0)}=0$. The corollary follows directly from Theorem \ref{Upperboundtheorem}.   
\Proofend

As a consequence of the lower and upper bound theorems  for $\{ \eta_{\ca'} \}$ (Theorems \ref{conjugate point theorem} and  \ref{Upperboundtheorem}), we have 3 linearly independent Jacobi fields, whose spatial part will not vanish nor diverge on the interval $[0,T]$. Hence their Euclidean norm with respect to the basis $\{e_\a \}$ is non-vanishing and bounded and lemma \ref{Equivalence lemma} implies that the two Euclidean norms are equivalent. $\{v, \eta_1, \eta_2, \eta_3 \}$ form a non-degenerate coordinate basis $\{\eta_\ca \}$ on $[0,T] $.

\noindent We now start the induction process.
\begin{assump}\label{TCurvassump}
In the proofs of the following lemmas we assume that 
\be
  \TD_v \TD^j_e \TCurv\tensor{\mu\nu}{K}{L} \, , \phm \TD^l_e \TCurv\tensor{\mu\nu}{K}{L} \quad \textrm{with} \phm  l \le j\le k+1 .
\ee
are bounded and that for $l \le j-1 $ these lemmas have been proven.
\end{assump}
\noindent As mentioned earlier, applying $\TD_v$ to (\ref{TCurvxitoe}) we deduce that
\be
   \TD_v \TD^j_\xi \TCurv\tensor{\mu\nu}{K}{L} \, , \phm \TD^l_\xi \TCurv\tensor{\mu\nu}{K}{L} \quad \textrm{with} \phm  l \le j,
\ee
are bounded as well. 

Closer analysis of (\ref{Qkformula}) shows that if we write all $\TCurv(v,\xi)\tensor{}{K}{L} $ as $\TCurv(v,e_\b)\tensor{}{K}{L} \xi^\b$, then $Q_{(j)}$ is made of derivatives of the tractor curvature and the tractor frame $\TE_\TB $ up to order $j$. The $y^\TB_{(l)}$, $l < j$, are bounded by assumption \ref{TCurvassump}. Hence the next step is to prove that the components of derivatives of $\TE_\TB $ are bounded.

\begin{lemma}{\textbf{[1,j]}\phm}
\label{DxiEk}
  $\TD^3_v \TD^{j}_\xi \TE^I_\TB$ and $\TD^{j}_\xi \TE^I_\TB $ are bounded.
\end{lemma}

\Proofstart Clearly this holds for $j=0$ and hence we start with $j=1$.

For $j=1$ we are given bounded $\norm\TCurv^{(1,1)} \norm$. Thus $\TD_v \TD_{e_\a} \TCurv $, $\TD_{e_\a} \TCurv  $ and $\TCurv $ are bounded. Since $\eta^\b$ and $\TD_v \eta^\b$ are bounded by proposition \ref{JT0}, we deduce that $\TD_v \TD_\xi \TCurv $, $\TD_\xi \TCurv  $ are bounded. Setting $e_\a=v$ or $\xi = v $ implies $ \TD^2_v \TCurv $ and $ \TD_v \TCurv  $ are bounded as well.

In the equation \be\label{mmm}\TD_v \TD_\xi \TE_\TB = \TCurv(v,e_\b)\TE_\TB \xi^\b+ \TD_\xi \TD_v \TE_\TB \ee the curvature term is bounded by $\norm\TCurv \norm\cdot \norm \xi \norm $. Since the third term vanishes for $A, E_\ap$, an application of Lemma \ref{integrateDY} gives bounded $\TD_\xi A$ and $\TD_\xi E_\ap$. Choosing $\TE_\TB=V$, the third term becomes $\TD_\xi A$. The right hand side is hence bounded and we apply lemma \ref{integrateDY} once more. For $\TE_\TB=Z$ the proof is analogous, but it is simpler to observe that $\TD_\xi Z^I= \xi^I $ is by definition a Jacobi tractor and $ \TD^3_v \TD_\xi Z^I $ is bounded by corollary \ref{JT0}.

Applying $\TD_v $ to (\ref{mmm})  we get $\TD_v \TCurv $, $\TD_v \xi^\b$ and $ \TD_v \TD_\xi \TE_\TB$. The first is bounded by our curvature assumption, the second by Corollary \ref{JT0} and the last has just been shown to be bounded. Hence $ \TD^2_v \TD_\xi \TE_\TB$ is bounded. As we are given $ \TD^2_v \TCurv$, we can repeat this once more to obtained that  $ \TD^3_v \TD_\xi \TE_\TB $ is bounded. 

For the level $j$ we recall first that $\TD_\xi Z$ is a Jacobi tractor and thus its $(3, j-1)$-derivatives are bounded by Proposition \ref{JTk} $[3,j-1]$.

\noindent For the other basis tractors we use (\ref{curvswapE}) to express $\TD_v \TD^j_\xi \TE_\TB$ in terms of curvatures and $ \TD^j_\xi \TD_v \TE_\TB$. Expanding $ \TD^l_\xi [\TCurv(v,\xi_{j-l})\TD^{j-(l+1)}_\xi \TE_\TB] $ gives contractions between  $\TD_v\TD^l_\xi \TCurv$, $ \Dhat_v \Dhat^l_\xi \xi $, $ \TD^{j-1}_\xi \TE_\TB$ and lower order terms. For $l < j$, all terms on the right hand side are bounded except for $ \TD^{j}_\xi \TD_v E_\TB$. But for $\TE_\TB = A, E_\ap$ the term vanishes hence proving that $\TD_v \TD^j_\xi A$ and $\TD_v \TD^j_\xi E_\ap$ are bounded. Now we can use this to prove the lemma for $V$. From Lemma \ref{integrateDY} we obtain that $\TD^{j}_\xi \TE_\TB $ are bounded.

Applying $\TD_v$ to (\ref{curvswapE}) and expanding, all terms on the right hand side are bounded by our curvature assumptions and lemma \ref{JTk} $[3,j-1]$. Thus the left side is bounded. Differentiating once more we find that the right hand side is bounded again, thus giving the desired boundedness of $\TD^3_v \TD^{j}_\xi \TE^I_\TB$ on the left hand side.
 \Proofend

\begin{lemma}{\textbf{[2,j]}}\label{Qk} 
 $Q_{(j)} $ and $\bar{Q}_{(j)}  $ are bounded.
\end{lemma}
\Proofstart We start with $j=1$. Thus in (\ref{Qkformula}) we have $\TD_\xi Q_{(0)} =0$, $Y_{(0)}=\JT$ and $y^\b_{(0)}=\eta^\b$. All the terms in $Q_{(1)}$ are contractions of $\TD_v \TD_\xi \TCurv  , \TD_\xi \TCurv $ and $\TCurv $ with $Y$, $\xi $ and the frame components $\eta^\b$ and $\xi $  as well as their first and second time derivative. We observe $\TD_\xi V^I = \TD_v \xi^I $. All these terms are bounded by Proposition \ref{JT0} or by Assumption \ref{TCurvassump}. Thus $Q_{(1)}$ is bounded.

Writing $\TD_\xi \JT = T + \bar{T} $ we find that $\bar{T}  $ contains terms $G(\JT, \TD_\xi \TE_\TB) $. $\bar{Q}_{(1)}$ is the third time derivative of $\bar{T}$ and can be decomposed into time derivatives of $\JT^\TB $ and $\TD_\xi \TE_\TB $ . Hence by corollary \ref{JT0} and lemma \ref{DxiEk}[1,1] we can conclude that $\bar{Q}_{(1)}$ is bounded. 

Now we treat the general case.  We start with $Q_{(j)}$. 
The only term of real interest is $\TD_\xi Q_{(j-1)} $, as the analysis of the other terms follows exactly the one of $Q_{(1)}$, with the observation that $Y^I$ and $y^\TB$ are of order $j-1$ and hence all necessary bounds are known. 
For $j \ge 2$, the term $\TD_\xi Q_{(j-1)} $ contains curvatures up to order $j+1$. The highest order terms will always contain at least one derivative along $v$ and at most $j$ along $\xi$. We use (\ref{curvswapE}) in reverse to pull the $\TD_v$-derivative to the front. This will generate extra curvature terms up to order $j-1$.
Since the our curvature bounds require the frame tractors to be contracted from the outside, we expand and apply (\ref{curvswapE}) again to pull them outside. We get at most derivatives of type $(1,j) $ or $(0,j) $ for the frame vectors and these have been bounded in lemma \ref{DxiEk}[1,j]. The remaining terms are all of lower order and thus bounded by earlier inductions steps. In particular for all $y^\c_{(l)}$ we must have $l<j$, as the terms of order $j$ have been isolated in the derivation of (\ref{generalconfJac}) and (\ref{Qkformula}). It follows that the tractor curvature bounds in assumption \ref{TCurvassump} are sufficient to prove the boundedness of $Q_{(j)}$. 

Recalling (\ref{defTTbar}) 
 and expanding $\bar{Q}_{(j)}$, we see that it contains terms of the form $\TD_v^q G(\TD_\xi^l\JT, \TD_\xi^m \TE_\TB)$ and $\TD_v^q\TD_\xi^m \TE^I_\TB$ for $q \le 3, \, l<j, \, m\le j $. By lemma \ref{JTk}[3,l] and \ref{DxiEk}[1,j] the terms $\TD_v^q \TD_\xi^l \JT^\TB $ and $\TD_v^q \TD_\xi \TE_\TB $ are bounded. This implies the boundedness of $\bar{Q}_{(j)}$ and hence the proof.
\Proofend

\begin{lemma}{\textbf{[3,j]}\phm} \label{JTk}
\newline
  $y^\TB_{(j)} = \TD^j_\xi (\JT^\TB)$, $Y=\TD^j_\xi \JT $ and their first three time derivatives are all bounded.
\end{lemma}
\Proofstart The case $j=0$ was already proven in corollary \ref{JT0}. 
All the conditions of the upper bound theorem are satisfied by assumption \ref{TCurvassump} and the lemmas \ref{DxiEk}, \ref{Qk}. Hence $y^\TB_{(j)} = \TD^j_\xi (\JT^\TB)$ is bounded. Using $Y^I=T^I+\bar{T}^I$ and expressing the right hand side in terms of $y^\TB_{(l)} $ and $\TD^m_\xi \TE_\TB$, all of which are bounded for $l,m \le j $, we deduce that $Y=\TD^j_\xi \JT$ is bounded. The same method provides a proof for the boundedness of $\TD^q_v \TD^j_\xi \JT$ for $q \le 3$.
\Proofend

We have shown that the lemmas hold for $j=0$ and $j=1$. Furthermore, under assumption \ref{TCurvassump} we can prove each lemma inductively from previous ones. Using assumption $(ii)$ of Theorem \ref{Maintheorem} we take $j=k+1$ and therefore deduce:
\begin{corollary}
\label{extensioncor}
There exists a $C^k$ extension for the metric $g_{ij}=\CV^{-2} \cg_{ij}$. Hence the conformal metric is regular on $U$.
\end{corollary}
\Proofstart
Recall that it is enough to show that the $(k+1)^{th} $ derivatives of the metric components (\ref{metricderivatives}) are 
 bounded on $\mathcal{O}$, which is Lemma \ref{JTk} $[3,k+1]$. The conditions of Whitney's Extension Theorem are therefore satisfied and the corollary follows.
\Proofend

We have proven part (ii) of Theorem \ref{Maintheorem} and shown the local extension of the conformal structure with $\cg$ and $\Tg$. 

We extend Lemma \ref{DxiEk} by the following corollary.
\begin{corollary}
\label{corEk}
  $\TD^2_v \TD^{k+1}_\xi \TE^I_\TB$ are bounded.
\end{corollary}
\Proofstart 
The proof is identical to that of Lemma \ref{DxiEk} for $k+1$. We use the fact that we have proven the boundedness of $y^\TB_{(k)} = \TD^k_\xi (\JT^\TB)$ by now and that (\ref{curvswapE}) requires tractor curvature up to order $k$
\Proofend

\subsection{Curvature bounds}

In this section we prove part (iii) of Theorem \ref{Maintheorem}, that the derivatives of the Riemann curvature of the general Weyl connection $\Dhat$ are bounded up to order $k$ in the conformally orthonormal frame $e_\a $ while those of the Riemann curvature of $g_{ij}$ are bounded up to order $k-1$.
This will be another induction, for $0\le j\le k$:
\bitem
\item Lemma \ref{DqPhatk} [4,$j$] shows that we get bounded derivatives of $\Phat(\xi,\eta) $ and $\langle b, \eta \rangle$ up to $j^{th}$ order.
\item Lemma \ref{Dxiek} [5,$j$] deduces the boundedness of $\Dhat^{j}_\xi e^i_\a$.
\item Lemma \ref{Dhatketa} [6,$j$] deduces the boundedness of $ \Dhat^{j}_\xi \eta^i_\ca$.
\item Lemma \ref{Rhatk} [7,$j$], proves that $\Dhat^{j}_\xi \Phat_{ik}, \Dhat^{j}_\xi C\tensor{ik}{l}{m}, \Dhat^{j}_\xi \Rhat\tensor{ik}{l}{m}$ are bounded.
\item Lemma \ref{Rk} [8,$j$], by showing that $ \Dhat^{j}_\xi b_i$ is bounded, we prove that the Riemann curvature associated to $g_{ij}$ is $C^{j-1}$.
\eitem
For the induction process, we make the following assumption.
\begin{assump}\label{Rhatassump}
For the following lemmas the conditions of Assumption \ref{TCurvassump}, and hence the Lemmas \ref{DxiEk}, \ref{Qk} and \ref{JTk}, are satisfied. Furthermore, we assume that the Lemmas \ref{DqPhatk}[4,l] - \ref{Rk}[8,l] have been proven for $l \le j-1 $.
\end{assump}

\begin{lemma}{\textbf{[4,j]}\phm}
\label{DqPhatk}   
  $\Dhat^{j}_\xi \Phat(\eta , e_\b) $ and $\Dhat^{j}_\xi \langle b, \eta \rangle $ are bounded.
\end{lemma}
\Proofstart For $j=0$, we observe $\Phat(\eta,e_\b) = E^I_\b \TD_\eta A_I = -A_I \TD_\eta E^I_\b $ is bounded by Corollary \ref{corEk}. 
To prove the boundedness of $ \langle b, \eta \rangle $, we use $\Phat(\eta , v) = \Dhat_v \langle b, \eta \rangle$ and integrate. Alternatively $A^I \JT_I = \langle b, \eta \rangle $ is bounded by corollary \ref{JT0}[3,0]. 

For the general proof we expand $\Dhat^{j}_\xi \Phat(\eta , e_\b) = \TD^{j}_\xi (E^I_\b \TD_\eta A_I)$. The terms in the expansion are of the form $ \TD^l_\xi \TE^I_\TB$, with $l \le j+1$. By Lemma \ref{DxiEk} and Corollary \ref{corEk} these are all bounded and hence  $\Dhat^{j}_\xi \Phat(\eta , e_\b) $ is. Using $e_\b = v $ gives the boundedness of  $\Dhat^{j}_\xi \langle b, \eta \rangle $ as before.
 \Proofend

The next lemma is the tensor version of Lemma \ref{DxiEk}. 
\begin{lemma}{\textbf{[5,j]}\phm}
\label{Dxiek}
  $\Dhat_v\Dhat^{j}_\xi e^i_\a$ and $\Dhat^{j}_\xi e^i_\a $ are bounded.
\end{lemma}
\Proofstart For $j=1$ we observe that the last term in $\Dhat_v \Dhat_\xi e^i_\a = \Rhat(v,\xi)\tensor{}{i}{j}e^j_\a + \Dhat_\xi \Dhat_v e^i_\a$ vanishes because the frame  $\{e_\a\} $ is Weyl propagated. The boundedness of $\Rhat\tensor{ik}{l}{m}$ implies that $\Dhat_v \Dhat_\xi e^i_\a$ is bounded. The lemma follows after application of lemma \ref{integrateDy}.

The general case applies the same ideas to (\ref{curvswape}), where assumption \ref{Rhatassump} assure the boundedness of the terms in the expansion. The last term vanishes as before and an application of Lemma \ref{integrateDy} concludes the proof.
\Proofend

\begin{lemma}{\textbf{[6,j]}\phm}
\label{Dhatketa}
$ \Dhat^j_\xi \eta^i $ are bounded
\end{lemma}
\Proofstart We recall that $\Tg(\JT, \TE_\b) = g(\eta,e_\b) = \eta_\b$. Expanding $\Dhat^j_\xi (\eta^\b e^i_\b) $ generates $\Dhat^l_\xi  e^i_\b $ and $\Dhat^l_\xi \eta^\b=\TD^l_\xi \JT_\b=y_{(l)}^\b$, where $l\le j$. These are bounded by Lemma \ref{Dxiek} and \ref{JTk}
\Proofend

\begin{lemma}{\textbf{[7,j]}\phm}
\label{Rhatk}
  We can obtain bounds with respect to $\{e_\a\} $ for the following curvature tensors: 
1)  $\Dhat^{j}_\xi \Phat_{ik}$,
2) $\Dhat^{j}_\xi C_{iklm}$,
3) $\Dhat^{j}_\xi \Rhat_{iklm}$.
\end{lemma}
\noindent Remark: Since the congruence has no conjugate points we can apply Lemma \ref{Equivalence lemma}.
\Proofstart 1) For $j=1$, this is simply a combination of Lemmas \ref{Equivalence lemma} and \ref{DqPhatk}[4,1].

\noindent For general $j$, we know from Lemma \ref{DqPhatk}[4,$j$] that $\Dhat^{j}_\xi \Phat(\eta, e_\b) $ is bounded. Expanding, gives $(\Dhat^{k}_\xi\Phat_{ik})\eta^i e^k_\b $ and derivatives of $\Phat_{ik} $, $e^i_\b $ and $\eta^i$, which are all bounded by the Lemmas \ref{Rhatk}[7,$l$], \ref{Dxiek}[5,$l$] and \ref{Dhatketa}[6,$l$] for $l \le j$. Hence $(\Dhat^{j}_\xi\Phat_{ik})\eta^i e^k_\b $ is bounded. The bound with respect to $\{e_\a\} $ follows from Lemma \ref{Equivalence lemma}. 

\noindent 2) For $j=0$, we set ${\mathcal{C}} =\c$ and ${\mathcal{D}}=\d$, so that $\TCurv_{\a\b\TC{\mathcal{D}}} = C_{\a\b\c\d}$ is clearly bounded.

\noindent For general $j$, we apply $\TD^{j}_\xi $ and expand the tractor parts on the left hand side. All individual terms are bounded by our curvature assumptions and Lemma \ref{DxiEk}. Therefore $\Dhat^{j}_\xi (C_{\a\b\c\d})$ is bounded. We expand again, isolating the term $ (\Dhat^{j}_\xi C)_{\a\b\c\d}$. All other terms are bounded by Lemmas \ref{Rhatk}[7,$l$] and \ref{Dxiek}[5,$m$], for $l < j, m \le j$ and hence the boundedness of $ (\Dhat^{j}_\xi C)_{\a\b\c\d}$ follows.

\noindent 3) All that remains is to use the curvature decomposition (\ref{curvdecomp}) and the fact that the conformal metric is covariantly constant to deduce that $(\Dhat^{j}_\xi \Rhat)_{\a\b\c\d}$ is bounded. \Proofend

We are also interested in $R\tensor{ij}{k}{l}$ the curvature of the unphysical space-time metric $g_{ij}= \CV^{-2}\cg_{ij}$. It is related to $\Rhat\tensor{ij}{k}{l}$ by
\be
  \Rhat\tensor{ij}{k}{l} - R\tensor{ij}{k}{l} = 2 S\tensor{l[j}{kq}{}\D_{i]}b_q + 2 b_q S\tensor{r[i}{kq}{}  S\tensor{j]l}{rp}{}  b_p \non
\ee
The tensor $S\tensor{ij}{kl}{}$ is covariantly constant for all general Weyl derivatives, so that we need to show that the 1-form $b_i$ has bounded derivatives up to order $j$. We recall that the 1-form $b_i$ satisfies $\Dhat_i \CV = b_i \CV$ and is the 1-form of the conformal geodesics in the $\CV$-gauge, the unit velocity gauge. Hence we have $\Dhat = \D + b $ and $\Dhat_k g_{ij}=-2b_k g_{ij}$. 

\begin{lemma}{\textbf{[8,j]}\phm}
\label{Rk}
  The 1-form $b_i$ has bounded $j^{th}$ derivatives. This implies that the curvature derivatives $\D^{j-1}_\xi R_{iklm}$ are bounded and the metric $g_{ik} $ has bounded $C^{j-1} $ curvature.
\end{lemma}
\Proofstart By Lemma \ref{DqPhatk}[4,$j$], $\Dhat^{j}_\xi \langle b, \eta \rangle $ is bounded.
For $l \le j-1$, the derivatives of $b_{i} $, $\eta^i $ and $e^i_\b$ are bounded by the Lemmas \ref{Rk}[8,$l$], \ref{Dxiek}[5,$l$] and \ref{Dhatketa}[6,$l$]. 
Thus expanding $\Dhat^{j}_\xi \langle b, \eta \rangle $, gives bounds for $(\Dhat^{j}_\xi b_i)\eta^i$. Applying Lemma \ref{Equivalence lemma} the bound with respect to $\{e_\a\} $ follows.
\Proofend

This lemma, for $j=k$, proves the final part, part (iii), of Theorem \ref{Maintheorem}. 
The following corollary will be useful in a later paper:
\begin{corollary}
  The kinematic quantities in the $\CV$-gauge are bounded.
\end{corollary}
\Proofstart $\Dhat_v \eta^i = \Dhat_\eta v^i = \D_\eta v^i + \langle b,\eta \rangle v^i +  \langle b,v \rangle \eta^i - g(v,\eta)b^i$ is bounded. It follows that $\D_\eta v^i$ and hence $\D_{e_\a} v^i$ are bounded. Since $v^i$ is a unit vector in the $\CV$-gauge, we can deduce that $\th, \s_{\a\b}, \om_{\a\b}$ are bounded. \Proofend

\subsection{Summary and discussion}

The three parts of Theorem \ref{Maintheorem} have been proven by Theorem \ref{conjugate point theorem}, Corollary \ref{extensioncor} and Lemma \ref{Rk}. 
The regular unphysical metric given by Theorem \ref{Maintheorem} is not unique, as a rescaling by $e^{f}$ with smooth $ f$ would give a regular metric too. Hence there are other conformal scales $\sigma $ and regular metrics $\sigma^{-2}\cg_{ij} $.


Our work incorporates the local extension theorem of R\'acz as stated in Theorem \ref{Racz theorem} above,  with time-like conformal geodesics replacing time-like  geodesics and conformally orthonormal frames replacing orthonormal ones, in the following sense: if we make the assumption of bounded (physical) curvature to order $k$ of Theorem \ref{Racz theorem} in our setup, then the tractor curvature is clearly bounded up to order $k-1$. Since in this case the Schouten tensor is finite, it follows that $\Om$ and $b$ will not diverge along the conformal geodesics. Therefore the local conformal extension of $g_{ij}$ is also a local extension for $g_{ij}$, as the conformal factor and its appropriate derivatives are extended analytically.  (R\'acz \cite{Racz} also gives an extension theorem based on null geodesics, and a corresponding result is subject of a separate paper \cite{Luebbenullextthm}.) 

Theorem \ref{Maintheorem} has two defects: firstly, that the extensions are only local i.e. the ideal endpoint of the incomplete conformal geodesic $\c $ may be the only point where the neighbourhood which is extended meets the boundary. The other curves in the congruence may be cut off by the selection of the neighbourhood $ \mathcal{{O}}$ without being incomplete at its boundary. Secondly, that the extension does not agree with the nonsingular parts of $M$ outside $U$. Taking a larger strongly causal neighbourhood of  $\mathcal{{O}}$ reaching all the way to the boundary, and attempting a simultaneous extension of all incomplete conformal geodesics in the congruence, two things could go wrong: the other conformal geodesics of the congruence might not be incomplete, reaching infinity instead; if they do end at the singularity then the tractor curvature could none-the-less blow up at their endpoints, as the boundedness assumption was only made in $\mathcal{{O}}$.

One would like to have a global extension theorem. The main problem with following our strategy is to show the existence of a congruence of incomplete conformal geodesics along which the tractor curvature is finite. One would need to analyse the Jacobi fields to show the absence of conjugate points, so that the coordinate system is well-defined. With strong enough assumptions, the procedure followed for the local extension does extend to this global problem, giving a global extension theorem of the following form:
\begin{theorem}
\label{global extension theorem}
Suppose we have a congruence $\Gamma$ of incomplete time-like conformal geodesics in $U \subseteq M$, with coordinates $\{ \s_\ca \} $ adapted to $\Gamma$, such that: 
\bitem 
\item All curves of $\Gamma$ end at the singularity $\Sigma \subset \partial U \subseteq \partial M$.
\item The norms of the spatial components of the Jacobi fields are bounded away from $0$ at $\Sigma$.
\item Strong causality and Whitney's property $ \mathcal{P}$ are satisfied in $U$.
\item The tractor curvature norms $\norm\TCurv^{(1,k+1)} \norm$ and $\norm\TCurv^{(k+1)}\norm $  of (\ref{curvnormfunctions}) are bounded in $U$.
\eitem
It follows that $\Sigma$ is a conformal gauge singularity and can be regularised by conformal rescaling.
\end{theorem}

These are strong assumptions. One would like to find curvature conditions sufficient for the construction of a congruence which automatically satisfied the first two properties. For perfect fluid Bianchi spacetimes it was observed in \cite{LuebbeTodBianchi} that the flow itself provides the desired congruence and a global extension theorem was proven using spatial homogeneity. 

A more general approach would be to ensure that our neighbourhood $ \mathcal{{O}}$ of $\c$ contains a cylinder of curves extending to the singularity. In other words, how can one be sure in the process of building $ \mathcal{{O}}$ in Proposition 3.2.6. of R{\'a}cz, that the 3-balls $\varsigma_t(\r) $ have a minimum radius $\r$?
Furthermore one would like to understand when the neighbourhood of a singularity satisfies the property $\mathcal{P}$. Given sufficient regularity of the singular set it should be possible to deduce $\mathcal{P}$ for certain space-times. 
These matters are under study.

\end{document}